\documentclass[aps, prd, groupedaddress, preprint, tightenlines, eqsecnum, nofootinbib, showpacs]{revtex4}
\usepackage{epsfig}
\usepackage{setspace}
\usepackage{graphicx}
\usepackage{leqno}
\usepackage{cancel}
\usepackage{amsmath}
\usepackage{amsfonts}
\usepackage{amssymb}
\usepackage{enumerate}

\usepackage{listings}
\usepackage{axodraw}
\usepackage{mathtools}

\usepackage{float, slashed, graphicx, amssymb, amsmath}
\usepackage[usenames, dvipsnames]{xcolor}
\usepackage{booktabs}

\def\Li{\mathop{\hbox{\rm Li}}\nolimits}

\def\PP{K}

\def\Psl{\slashed{K}}

\def\Fcc{F^{cc}}
\def\Fcc{P^{(2)}}

\def\spa#1.#2{\left\langle#1\,#2\right\rangle}
\def\spb#1.#2{\left[#1\,#2\right]}
\def\la{\langle}
\def\ra{\rangle}

\def\Atree{A^{\tree}}
\def\Aloop{A^{\oneloop}}

\def\ksl{\slashed{k}}

\newcommand{\oneloop}{\text{1-loop}}

\newcommand{\tree}{\text{tree}}

\DeclareMathOperator{\tr}{\mathrm{tr}}

\newcommand\figref[1]{fig.~\ref{#1}}
\def\eps{\epsilon}

\def\be{\begin{equation}}
\def\ee{\end{equation}}

\def\Log{{\rm Log}}

\begin{document}

\title{Analytic all-plus-helicity gluon amplitudes in QCD}

\author{David~C.~Dunbar, John~H.~Godwin, Guy~R.~Jehu and Warren~B.~Perkins}

\affiliation{
College of Science, \\
Swansea University, \\
Swansea, SA2 8PP, UK\\
\today
}

\begin{abstract}

We detail the calculation of two-loop  all-plus helicity amplitudes for pure Yang-Mills theory.
The four dimensional unitarity methods and augmented recursion techniques we have developed, together with a knowledge of the singular structure of the amplitudes
allow us to compute these
in compact analytic forms.
Specifically we present the computation and analytic results for the six- and seven-gluon leading colour two-loop amplitudes these being the first QCD two-loop amplitudes beyond five-points. 

\end{abstract}


\maketitle

\section{Introduction}

Perturbative scattering amplitudes are a key element in confronting particle theories with experiment. In doing so there is a need for precise theoretical 
predictions which require matrix elements beyond the ''leading order'' (LO)  and for many processing beyond 
''next-to-leading order'' (NLO)~\cite{Badger:2016bpw,Heinrich:2017una}.
These scattering amplitudes are challenging to compute due to both the proliferation of 
diagrams and the complexity of the associated integrals.

Enormous progress has been made in computing scattering amplitudes in maximally supersymmetric Yang-Mills using the enhanced symmetry together with calculations based upon the singular structure of the amplitude (for example \cite{Caron-Huot:2016owq}). 
However it is an open question as to whether these techniques can be used in non-supersymmetric theories.  
These methods utilise a knowledge of the factorisation properties of the amplitude~\cite{Eden,Britto:2005fq};
the unitarity of the $S$-matrix and its associated cut singularities~\cite{Bern:1994zx,Bern:1994cg}; and the general singular behaviour in the
''Infra-Red'' (IR) and ''Ultra-Violet'' (UV)~\cite{Catani:1998bh}. 

In this article we demonstrate the computation of a very specific two-loop amplitude in pure Yang-Mills theory where the external gluons all have the same helicity 
(all-plus) and we consider the leading in colour  component.  For two-loop QCD the four-point amplitude has been computed analytically for all helicity 
configurations~\cite{Glover:2001af,Bern:2002tk} and numerically using unitary techniques~\cite{Abreu:2017xsl}.
Recently, the first five-point two-loop amplitude for the all-plus helicity configuration   was obtained using 
$D$-dimensional unitarity\footnote{where $D=4-2\epsilon$.} to generate the integrands of loop integral functions~\cite{Badger:2013gxa,Gehrmann:2015bfy}. 
In \cite{Dunbar:2016aux} the amplitude
was recomputed using four dimensional unitarity techniques together with factorisation properties of the amplitude rather than the integrand.  
In \cite{Dunbar:2016gjb} we presented the result of applying this approach to computing the six-point all-plus helicity amplitude. In this article 
we detail this construction and present both the six- and seven-point all-plus helicity amplitudes. The expression for these are compact analytic formulae. 

\section{The all-plus helicity amplitude}

First we review the  all-plus helicity amplitude.
The all-plus helicity amplitude at leading colour may be written\footnote{The factor $c_{\Gamma}$
is defined as $\Gamma(1+\epsilon)\Gamma^2(1-\epsilon)/\Gamma(1-2\epsilon)/(4\pi)^{2-\epsilon}$.} 
\begin{align}
{\mathcal A}_{n}(1^+, 2^+,..., n^+) |_{\rm leading \; color}=& g^{n-2}  \sum_{L \ge 1} \left( g^2 N_c c_{\Gamma}\right)^L  
\times \sum_{\sigma \in S_{n}/Z_{n}} {\rm tr}(T^{a_{\sigma(1)}} T^{a_{\sigma(2)}} 
T^{a_{\sigma(3)}} \cdots  T^{a_{\sigma(n)}}) 
\notag\\ &
\times A^{(L)}_{n}(\sigma(1)^{+},  \sigma(2)^{+} ,..., \sigma(n)^{+})\,
\label{eq:fullamp}
\end{align}
where $N_c$ is the number of colours and the summation is over the set of non-cyclic permutations, $S_n/Z_n$. 

The tree amplitude vanishes as a consequence of supersymmetric Ward identities so the leading contribution is the one-loop amplitude
which, is given by~\cite{Bern:1993qk}\footnote{As usual,  a null momentum is represented as a
pair of two component spinors $p^\mu =\sigma^\mu_{\alpha\dot\alpha}
\lambda^{\alpha}\bar\lambda^{\dot\alpha}$. For real momenta
$\lambda=\pm\bar\lambda^*$ but for complex momenta $\lambda$ and
$\bar\lambda$ are independent~\cite{Witten:2003nn}.  
We are using a spinor helicity formalism with the usual
spinor products  $\spa{a}.{b}=\epsilon_{\alpha\beta}
\lambda_a^\alpha \lambda_b^{\beta}$  and 
 $\spb{a}.{b}=-\epsilon_{\dot\alpha\dot\beta} \bar\lambda_a^{\dot\alpha} \bar\lambda_b^{\dot\beta}$. Also
 $ s_{ab}=(k_a+k_b)^2=\spa{a}.b \spb{b}.a$.}, 
\begin{align}
A^{(1)}_n(1^+,2^+,\cdots,n^+)=-{i\over 3}\sum_{1\leq k_1<k_2<k_3<k_4\leq n} 
{\spa{k_1}.{k_2} \spb{k_2}.{k_3}\spa{k_3}.{k_4}\spb{k_4}.{k_1} \over \spa{1}.2\spa{2}.3 \cdots\spa{n}.1}  
+O(\epsilon) \; . 
\end{align}
This expression is rational to order $\epsilon^{0}$.  
All-$\epsilon$ forms of the one-loop amplitudes are given in terms of higher dimensional scalar integrals 
and for $n\leq 6$ are~\cite{Bern:1996ja}
$$
A_{4}^{(1)}(1^+,2^+,3^+,4^+) =
{ 2i \eps(1-\eps)\over  \spa1.2\spa2.3\spa3.4\spa4.1}
\times s_{12}s_{23} I_4^{D=8-2\eps} \,,
$$
\begin{eqnarray}
A_{5}^{(1)}(1^+,2^+,3^+,4^+,5^+) &=&
{ i \eps(1-\eps)\over  \spa1.2\spa2.3\spa3.4\spa4.5\spa5.1}
\notag \\
&\times &
\Bigl[
s_{23}s_{34} I_4^{(1),D=8-2\eps}
+s_{34}s_{45} I_4^{(2),D=8-2\eps}
+s_{45}s_{51} I_4^{(3),D=8-2\eps}
\notag \\
&+ & s_{51}s_{12} I_4^{(4),D=8-2\eps}
+s_{12}s_{23} I_4^{(5),D=8-2\eps}
+(4-2\eps) { \varepsilon (1,2,3,4) }I_5^{D=10-2\eps}
\Bigr] \,,
\notag
\end{eqnarray}
\begin{eqnarray}
 A_{6}^{(1)}(1^+, 2^+, 3^+, &   4^+ & , 5^+, 6^+) =
 {i \eps(1-\eps)\over \spa1.2 \spa2.3 \spa3.4 \spa4.5 \spa 5.6 \spa6.1}
 \frac{1}{2} \biggl[
  \notag \\
&  & 
- \hskip -.3 cm \sum_{1\leq i_1<i_2 \leq 6} \hskip -.2 cm
\tr[\ksl_{i_1} \Psl_{i_1+1, i_2-1}
\ksl_{i_2} \Psl_{i_2+1,i_1-1}] I_{4}^{(i_1,i_2),D=8-2\eps}
+  (4-2\eps)\, \tr[123456] \, I_6^{D=10-2\eps} \notag \notag \\
&  & \hskip 2 cm
+  (4-2\eps) \sum_{i=1}^6 \varepsilon(i+1, i+2, i+3, i+4)
I_{5}^{(i),D=10-2\eps} 
\biggr] \,, 
\end{eqnarray}
where $I_m^{(i),D}$ denotes the $D$ dimensional scalar integral obtained by removing the
loop propagator between legs $i-1$ and $i$ from the $(m+1)$-point
scalar integral etc.~\cite{Bern:1993kr},  $\Psl_{a,b}\equiv \sum_{i=a}^b \slashed{k}_i$ and
$\varepsilon(a,b,c,d)=\spb{a}.b\spa{b}.c\spb{c}.d\spa{d}.a-\spa{a}.b\spb{b}.c\spa{c}.d\spb{d}.a$.

The subject of this computation is the two-loop partial amplitude $A^{(2)}_n(1^+,2^+,...,n^+)$. 
This particular partial amplitude has full cyclic symmetry and flip symmetry, 
\begin{equation}
A^{(2)}_n(1^+,2^+,...,n^+) = (-1)^n A^{(2)}_n(n^+,...,2^+,1^+)
\; .
\end{equation}

The IR and UV behaviours of this amplitude are known~\cite{Catani:1998bh} and motivate a partition of the amplitude:
\begin{align}\label{definitionremainder}
A^{(2)}_{n}(1^+, 2^+,..., n^+)=& A^{(1)}_{n}(1^+, 2^+,..., n^+)
\left[ - \sum_{i=1}^{n} \frac{1}{\epsilon^2} \left(\frac{\mu^2}{-s_{i,i+1}}\right)^{\epsilon} 
+\frac{n\pi^2}{12} \right] +  \;F^{(2)}_{n}  + {\mathcal O}(\eps)\, .
\end{align}
In this equation $A^{(1)}_{n}$ is the all-$\epsilon$ form of 
the one-loop amplitude. 
The finite remainder function $F_n^{(2)}$ can be split into polylogarithmic and rational pieces,
\begin{equation}
F_n^{(2)} = \Fcc_n+R_n^{(2)}\; .
\end{equation}
For this amplitude $\Fcc_n$ can be computed using four-dimensional cuts . The result for arbitrary $n$ was calculated in \cite{Dunbar:2016cxp}
and for completeness we summarize the calculation in the following section. 
$R_n^{(2)}$ is a purely rational function whose structure is not captured by four-dimensional unitarity cuts. 
The result for $R_6^{(2)}$ was presented in \cite{Dunbar:2016gjb}.  Here the six-point example is used to present the methodology of the calculation. 
This methodology can be straightforwardly applied to the seven-point case and we present the result for $R^{(2)}_7$.

\section{Determining $\Fcc_n$}

In this section we review the construction of the pieces of the amplitude which contribute to the IR singular terms and 
 $\Fcc_n$. We are able to obtain these using four-dimensional unitarity methods. 
 Noting that the one-loop amplitudes have no four dimensional cuts since the order $\epsilon^0$ expression is purely rational, 
by using cuts where the momenta lie in four dimensions the one-loop  all-plus amplitude appears as a rational vertex which cannot be further cut. 
Consequently the cuts of the two-loop amplitude manifest as  cuts of one-loop integral functions which can be evaluated using essentially one-loop unitarity techniques.  For completeness we review the $n$-point calculation presented in~\cite{Dunbar:2016cxp}.

\begin{figure}[h]
\includegraphics{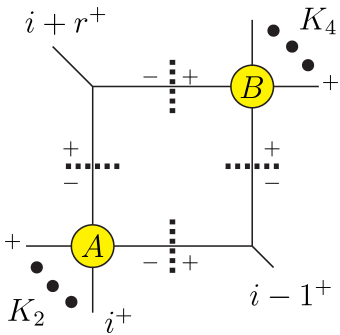}
    \caption{The non-vanishing quadruple cut. $A$ is a MHV tree amplitude whereas $B$ is a one-loop all-plus amplitude. The set $K_2$ may 
    consist of a single leg but the set $K_4$ must contain at least two legs. The integral function depends upon $S\equiv (k_{i-1}+K_2)^2$ and
    $T\equiv (K_2+k_{i+r})^2$.}
    \label{fig:oneloopstyle}
\end{figure}    
First consider quadruple cuts of the amplitude~\cite{Britto:2004nc}:  the only non-vanishing configuration is
 shown in fig.~\ref{fig:oneloopstyle}.
As shown in~\cite{Dunbar:2016cxp}, the coefficient of the one-loop box integral function $I_4^{2m}$ depicted in \figref{fig:oneloopstyle} is 
\begin{align}
&{ [i-1|K_4|i+r\ra 
{ [i+r|K_4|i-1\ra }    \over 
\spa1.2\spa2.3\spa3.4\cdots \spa{n}.1   }
\notag \\
& \hskip 2truecm\times
\left( 
\sum_{a<b<c<d \in K_4} \tr_{-}[abcd]-\sum_{a<b<c\in K_4} \tr_-[abc K_4]  +\sum_{a<b\in K_4} { \la i-1 |K_4  a b   K_4 | i+r\ra  \over \spa{i-1}.{i+r} }
\right) \; 
\end{align}
where
\begin{align}
\tr_{-} [abcd] \equiv \spa{a}.b\spb{b}.c\spa{c}.d\spb{d}.a \; ,
\end{align}
$K_4$ is the sum of the momenta in the set  $\{ i+r+1, \cdots , i-2\}$ with a cyclic definition of indices and inequality refers to ordering within the set.
The corresponding integral function satisfies
\begin{eqnarray}
-2 (ST-K_2^2K_4^2) I_4^{2m}
=   & & \left(  -{ (-S/\mu^2)^{-\epsilon} \over \eps^2 }
  -{ (-T/\mu^2)^{-\epsilon} \over \eps^2 }
+{ (-K_2^2/\mu^2)^{-\epsilon} \over \eps^2 }
+{ (-K_4^2/\mu^2)^{-\epsilon} \over \eps^2 } \right)
\notag \\
 & & +F^{2m}[S,T,K_2^2,K_4^2] 
\end{eqnarray}  
where  
\begin{align}
F^{2m}[ S,T, K_2^2, K_4^2] = 
&\Li_2[1-\frac{K_2^2}{S}]+
\Li_2[1-\frac{K_2^2}{T}]+
\Li_2[1-\frac{K_4^2}{S}]
\notag \\
+&
\Li_2[1-\frac{K_4^2}{T}]-
\Li_2[1-\frac{K_2^2K_4^2}{ST}]+
\Log^2(S/T)/2 \; . 
\end{align}

There are also contributions from one and two mass triangle integral functions.
These can be determined using triple cuts 
(for example using canonical forms~\cite{Dunbar:2009ax} or analytic structure~\cite{Forde:2007mi}) and only contribute to the IR singularities. 
Overall~\cite{Dunbar:2016cxp}, 
\begin{align}&
\sum  {\cal C}_{i} I_{4,i}^{\rm 2m} 
\biggl|_{IR}
+\sum  {\cal C}_{i}  I_{3,i}^{2 \rm m}  
+\sum  {\cal C}_{i}  I_{3,i}^{1 \rm m} 
= 
A^{(1),\epsilon^0}_n(1^+,2^+,\cdots, n^+)
\times -\sum_{i=1}^{n} \frac{1}{\epsilon^2} \left(\frac{\mu^2}{-s_{i,i+1}}\right)^{\epsilon} ,
\end{align}
where $A^{(1),\epsilon^0}_n(1^+,2^+,\cdots, n^+)$ is the order $\epsilon^0$ truncation of the one-loop amplitude. 

A key step is to promote the coefficient of these terms to the all-$\epsilon$ form of the one-loop amplitude. This ensures that the two-loop amplitude
has the correct singular structure.

$P_n^{(2)}$ is obtained by summing over all possible box contributions, including the degenerate cases when $K_2$  corresponds to a single leg 
($K_4$ must contain to at least two external legs).  
The full expression for $\Fcc_n$ is~\cite{Dunbar:2016cxp}, 
\begin{equation}
P_n^{(2)}   =  -{ i  \over 
3 \spa1.2\spa2.3\spa3.4\cdots \spa{n}.1   }\sum_{i=1}^n  \sum_{r=1}^{n-4} c_{r,i}  F^{2m}_{n:r,i}
\end{equation}
where the labels on the integral function correspond to those in  \figref{fig:oneloopstyle} and the coefficient is 
\begin{align}
c_{r,i}= \left( 
\sum_{a<b<c<d \in K_4}  \tr_{-} [abcd]-\sum_{a<b<c \in K_4} \tr_{-}[abc  K_4]  +\sum_{a<b \in K_4 } { \la {i-1} |K_4  a b K_4 | {i+r} \ra  \over \spa{{i-1}}.{{i+r}} }
\right)\; 
\end{align}

\def\ki{i}
\def\kj{j}
\def\kk{k}
\section{Complex Recursion}

Britto-Cachazo-Feng-Witten recursion \cite{Britto:2005fq} exploits the analytic properties of a rational quantity, $R$, under
a complex shift of its external momenta. $R$ may be an $n$-point tree amplitude or part of a loop amplitude. The momentum shift
introduces a complex parameter, $z$, whilst  preserving  overall momentum conservation and keeping all external momenta null.  

Possible shifts include the original BCFW shift which acts on two momenta, say  $p_\ki$ and $p_\kj$, by
\begin{equation}
\bar\lambda_{{\ki}}\to \bar\lambda_{\hat{\ki}} =\bar\lambda_\ki - z \bar\lambda_\kj 
\qquad , \qquad
\lambda_{{\kj}}\to\lambda_{\hat{\kj}} =\lambda_\kj + z \lambda_\ki \; ,
\label{BCFWshift}
\end{equation}
the all-line shift~\cite{Cohen:2010mi}
and
the Risager shift~\cite{Risager:2005vk} which acts on three momenta, say $p_\ki$, $p_\kj$ and $p_\kk$, by
\begin{align}
\lambda_\ki\to &\lambda_{\hat \ki} = \lambda_\ki\,\, +z\, [\kj\kk] \lambda_\eta \,,
\notag \\
\lambda_\kj\to &\lambda_{\hat \kj} = \lambda_\kj\, +z\spb{\kk}.\ki \lambda_\eta \,,
\notag \\
\lambda_\kk\to &\lambda_{\hat \kk} = \lambda_\kk +z\spb{\ki}.\kj \lambda_\eta \;.
\label{KasperShift}
\end{align}
In the last case $\lambda_\eta$ must satisfy $\spa{\ki}.{\eta}\neq 0$ etc., but is otherwise unconstrained.

After applying the shift, the rational quantity of interest is a complex function parametrized by $z$ i.e. $R(z)$. Cauchy's theorem implies that:
\begin{equation}
\label{Amp AsResidues}
{1\over 2\pi i}\oint {R(z)\over z}= R(0) +\sum_{z_j\neq 0} {\rm Res}\Bigl[ {R(z)\over z}\Bigr]\Bigr\vert_{z_j}\; ,
\end{equation}
where the contour is a circle taken towards infinity. If $R(z)$ vanishes in the large $\vert z\vert$ limit, the left hand side is zero and the 
unshifted quantity  can be defined in its entirety by its singularities:
\begin{equation}
R= R(0)= -\sum_{z_j\neq 0} {\rm Res}\Bigl[ {R(z)\over z}\Bigr]\Bigr\vert_{z_j}\; .
 \label{Amp AsResidues2}
\end{equation}

For the rational part of the two-loop all-plus amplitude the 
BCFW shift generates a shifted quantity that does not vanish at infinity and so cannot be used to reconstruct the amplitude (the one-loop all-plus amplitudes
also behave in this way). However, using the Risager shift~\eqref{KasperShift} does yield a shifted quantity with the desired asymptotic behaviour, so this
is the shift employed.

Eq.~\eqref{Amp AsResidues2} is true when $R(z)$ has either simple or higher order poles.  In the case of a simple pole at 
$z=z_j$,
\begin{equation}
{\rm Res}\Bigl[ {R(z)\over z}\Bigr]\Bigr\vert_{z_j}\; =\frac{1}{z_j} {\rm Res}\Bigl[ {R(z)}\Bigr]\Bigr\vert_{z_j}
\end{equation}
and an understanding of the factorisations of $R(z)$ can be used to obtain the residue. For example, tree amplitudes have simple poles when a 
shifted propagator vanishes and the corresponding residues  are readily obtained from general factorisation theorems
leading to the BCFW recursion formulae for tree amplitudes~\cite{Britto:2005fq}.

Loop amplitudes in non-supersymmetric theories may have double poles. Mathematically this is not a problem since if we consider a function with a 
double pole at $z=z_j$  and 
Laurent expansion,
\begin{eqnarray}
R(z) &=& \frac{c_{-2}}{(z-z_{j})^2}+ \frac{c_{-1}}{(z-z_{j})}+\mathcal{O}((z-z_{j})^0)\; , \notag \\
\end{eqnarray}
then the required residue is 
\begin{equation}
{\rm Res}\Bigl[ {R(z)\over z}\Bigr]\Bigr\vert_{z_j}  = 
-\frac{c_{-2}}{z_{j}^2}+ \frac{c_{-1}}{z_{j}}
\end{equation}
and we can use Cauchy's theorem {\it provided} we know the value of both the leading and sub-leading poles. 
When applying this to an amplitude, the leading pole can be obtained from factorisation theorems, but,
at this point, there are no general theorems determining the sub-leading pole. Consequently we need to determine the 
sub-leading pole for each specific case. 
We do this by identifying the singularities within a specific computational 
scheme, we choose to use axial gauge. We have used this approach previously to compute one-loop 
which contain double poles~\cite{Dunbar:2010xk,Alston:2015gea,Dunbar:2016dgg}.
 We label this process  {\em augmented recursion} and  describe it in later sections.

\section{Factorisations of $R_n^{(2)}$}

To apply the  methods of the previous section we must discuss the factorisations and resultant residues of $R_n^{(2)}$. 

Note that $P_n^{(2)}$ does not contain any unphysical poles and so the factorisations of $R_n^{(2)}$ are the physical factorisations of the amplitude. 
Explicitly these are
\begin{align}
A^{(1)}_{m}
(a^+,\cdots, b^+,\hat{K}^\pm)&{1 \over K_{a,b}^2 }  A^{(1)}_{n+2-m}(-\hat{K}^{\mp},b+1^+,\cdots, a-1^+ )
\shortintertext{with $3\leq m\leq n-1$ and}
R^{(2)}_{n-1}(a+2^+,\cdots a-1^+, \hat{K}^{+})&{1\over s_{a\, a+1}}A^{(0)}_{3}({a}^+,a+1^+,-\hat{K}^-)\;
\end{align}
where the particular one-loop amplitudes $A^{(1)}_r$ in the factorisation are purely rational functions.

Many factorisations of loop amplitudes involve only simple poles and the corresponding residues can be determined using general factorisation theorems.
However certain factorisations  introduce double poles which are sensitive to sub-leading information that is not captured by the general 
theorems.

\begin{figure}[H]
\begin{center}
\includegraphics{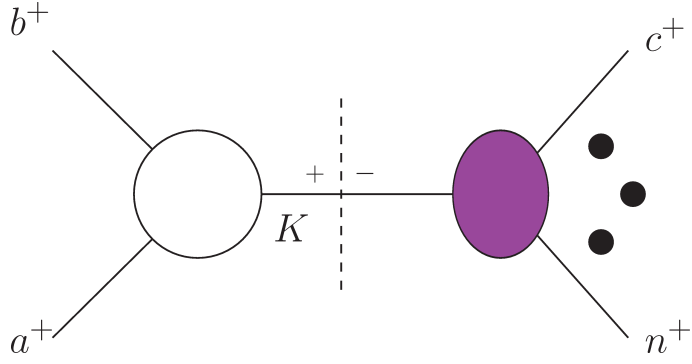}
\caption{The origin of the double pole. The double pole corresponds to the coincidence of the singularity arising in the 3-pt  
integral with the factorisation corresponding to $K^2=s_{ab}\to 0$.}
\label{fig:doublepole}
\end{center}
\end{figure}

The origin of these terms is shown in fig.~\ref{fig:doublepole} 
where there is an explicit pole from the highlighted propagator and a further pole from the 3-pt loop integral. 

\section{Augmented Recursion}

To analyse the $s_{ab}$ double pole we use axial gauge methods.
The advantage of axial gauge is that, although it is an off-shell method, helicity still labels the internal legs and the vertices 
can be expressed 
in terms of {\em nullified momenta}~\cite{Schwinn:2005pi,Kosower:1989xy}  defined using,
\begin{align}
K^\flat = K - \frac{K^2}{[q|K|q\ra}q \nonumber \\
\label{eq:nullified}
\end{align}
where $q$ is a reference vector. 

\begin{figure}[H]
\begin{center}
\includegraphics{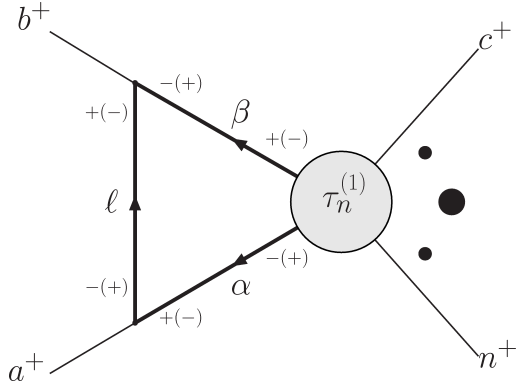}
\end{center}
\caption{Diagram containing the leading and sub-leading poles as $s_{ab}\to 0$. The axial gauge construction permits the off-shell continuation of the internal 
legs. The two 
internal helicity configurations must be summed 
over to obtain the complete contribution to the $1/s_{ab}$ residue. The one-loop current $\tau_n^{(1)}$ can be built from the on-shell $n$-point one-loop 
single minus amplitude.}
\label{fig:axialg}
\end{figure}

In this formalism the 3-pt loop integral can be expressed as, 
\begin{equation}
V_3^{(1)}(a^+,b^+,K^+)= \frac{i}{3}\frac{\spb a.b\spb b.{K^{\flat}} \spb{K^{\flat} }.a}{K^2}
\end{equation}
which is only non-zero for complex momenta. 
The pole as $K^2\to 0$ arises from regions of the loop momentum integration where three adjacent propagators are simultaneously on-shell.  
The full contribution arises from structures of the form shown in fig.~\ref{fig:axialg} which contain both single and double poles.

\begin{figure}[H]
\begin{center}
\includegraphics{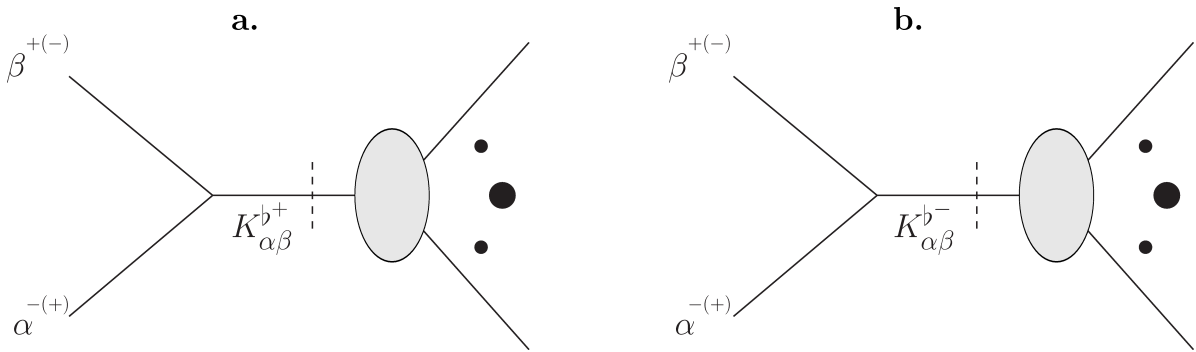}
\end{center}
\caption{The factorisations that $\tau_n^{(1)}$ must reproduce.}
\label{fig:doublepoleB}
\end{figure}

Expressing the 3-pt vertices in axial gauge,  the principal helicity assignment in \figref{fig:axialg} gives the integral
\begin{align}
 \frac{i}{(2\pi)^D}\int\!\! \frac{d^D\ell}{\ell^2\alpha^2\beta^2} 
\frac{[a|\ell|q\ra[b|\ell|q\ra }{\spa a.q\spa b.q} \frac{\spa \beta.q^2}{\spa\alpha.q^2}\tau_n^{(1)} (\alpha^{-},\beta^{+},c^+,...,n^+) \; .
\label{eq:tauintinitA}
\end{align}
To determine \eqref{eq:tauintinitA} in general we would need to consider $\ell$, $\alpha$  and $\beta$ to be off-shell and $\tau_n^{(1)}$ to be the 
doubly off-shell one-loop current. 
However, as we are only interested in the residue on the $s_{ab}\to 0$ pole, we do not need the full current.
Instead it is sufficient for  $\tau_n^{(1)}$ to satisfy two conditions \cite{Dunbar:2016dgg}:

\begin{itemize}
\item[\textbf{I}] 
it reproduces the leading poles in  $s_{\alpha\beta} \rightarrow 0$ shown in fig.~\ref{fig:doublepoleB}.

\item[\textbf{II}] in the limit $\alpha^2, \beta^2 \rightarrow 0$  (the on-shell limit) the $n$-point one-loop single minus amplitude is reproduced.
\end{itemize}

In manipulating $\tau_n^{(1)}$ we can neglect terms which are ultimately finite as $\spa a.b\to 0$ and hence do not contribute to the residue.
As any powers of $\ell^2$, $\alpha^2$ or $\beta^2$ appearing in the numerator generate powers of $s_{ab}$ when the loop momentum integration is performed,   
 $\alpha^2$ and $\beta^2$ are also regarded as small in these manipulations. 
 Similarly $\spa a.\alpha$, $\spa a.\beta$, $\spa b.\alpha$ and 
 $\spa b.\beta$ can all be regarded as small.
 
Condition \textbf{II} provides a starting point for constructing the current as detailed in appendix~\ref{app:sixpoint}.

\section{The $s_{ab}$ pole in the Six-Point Rational Term}

In this section we determine the  leading and sub-leading poles in $R_6^{(2)}$ as $s_{ab}\to 0$. 
Our starting point is the 
 on-shell six-point single minus amplitude organised as in appendix~\ref{app:sixpoint}:
\begin{equation}
A_6^{(1)}(\alpha^-,\beta^+,c^+,d^+,e^+,f^+)= \tau_6^{\rm d.p}+\tau_6^{\rm s.b}+\tau_6^{\rm n.f}+{\cal O}(\spa{\alpha}.{\beta})
\end{equation}
which we simply take off-shell by replacing $\alpha$ and $\beta$ by their nullified form within axial gauge.
With this organisation it is natural to write
\begin{align}
 \frac{i}{c_\Gamma(2\pi)^D}\int\!\! \frac{d^D\ell}{\ell^2\alpha^2\beta^2} 
\frac{[a|\ell|q\ra[b|\ell|q\ra }{\spa a.q\spa b.q} \frac{\spa \beta.q^2}{\spa\alpha.q^2}\tau_6^{(1)} (\alpha^{-},\beta^{+},c^+,d^+,e^+,f^+) 
=  \mathcal{I}^{\alpha-\beta+}_{\text{d.p.}}+\mathcal{I}^{\alpha-\beta+}_{\text{s.b.}}+\mathcal{I}^{\alpha-\beta+}_{\text{n.f.}} \;.\label{eq:tauintinitB}
\end{align}

The double poles arise from
\begin{align}
\mathcal{I}^{\alpha-\beta+}_{\text{d.p.}} &\equiv \frac{i}{3}\int\!\! d\Lambda^{\alpha\beta}_{ab} 
\Biggl[\frac{\la q|\alpha\beta|q\ra[q|\PP_{ab} |q\ra}{s_{ab}[q|\PP_{ab} |c\ra\spa{d}.e^2 }\Biggr( -{[fc]^3\over t_{abc}[f|\PP_{ab}|c\ra}\frac{[q|\PP_{ab}|c\ra^3}{[q|\PP_{ab}|q\ra^3}
\notag \\ & \quad\quad\quad\quad\quad\quad\quad\quad\quad\quad\quad +{ [ef]\spa{d}.f \over\spa{c}.d\spa{e}.f^2}\frac{[q|\PP_{ab}|e\ra^3}{[q|\PP_{ab}|q\ra^3}
-{\spa{c}.e [dc]\over\spa{c}.d^2\spa{e}.f }\frac{[q|\PP_{ab}|d\ra^3}{[q|\PP_{ab}|q\ra^3}\frac{[q|\PP_{ab}|c\ra}{[q|\PP_{ab}|f\ra }\Biggr)\Biggr]
\end{align}
where\footnote{The factor $c_\Gamma$ is inserted to make the normalisation consistent with eq.~\ref{eq:fullamp}.}
\begin{equation}
\int\!\! d\Lambda^{\alpha\beta}_{ab}  \equiv \frac{i}{c_\Gamma(2\pi)^D}\int\!\! \frac{d^D\ell}{\ell^2\alpha^2\beta^2} 
\frac{[a|\ell|q\ra[b|\ell|q\ra }{\spa a.q\spa b.q}  \;.
\end{equation}

The integral corresponding to the square bracket pole in the current is,
\begin{align}
\mathcal{I}^{\alpha-\beta+}_{\text{s.b.}}=& \frac{i}{3}\int\!\! d\Lambda^{\alpha\beta}_{ab}\Biggr[{[q|\beta|q \ra ^2\over [q|\alpha|q\ra^2}{[q|\alpha\beta|q]\over  [q|\PP_{ab}|q\ra^2 }{1\over s_{ab}}\frac{[f|\PP_{ab}|q\ra^2}{t_{cde}}\biggl( {\la q |\PP_{ab}|c][cd]\over \spa{d}.e\la q |\PP_{ab}\PP_{cd}|e\ra}
\notag \\ &
\quad\quad\quad\quad\quad\quad\quad\quad\quad\quad\quad\quad\quad\quad\quad\quad\quad\quad\quad\quad -{[de][ef]\over \spa{c}.d[f|k|c\ra}+{[ce]\over\spa{c}.d\spa{d}.e} \biggr)\Biggr]\; .
\end{align}
The remaining function constitutes the sub-leading non-factorising parts: 
\begin{equation}
\mathcal{I}^{\alpha-\beta+}_{\text{n.f.}} =
\int\!\! d\Lambda^{\alpha\beta}_{ab} \Biggr[\frac{\spa \beta.q^2}{\spa \alpha.q^2} \times  \tau_6^{\rm n.f}\Biggr]\quad ,
\end{equation}
where $\tau_6^{\rm n.f}$ is the expression defined in (\ref{eq:taunf}).

These expressions may be integrated to obtain the contribution from \figref{fig:axialg} (up to terms that are finite as $s_{ab}\to 0$) . The explicit integrations may be found in appendix~\ref{app:B}.
The results are\footnote{The restriction $|_{\mathbb{Q}}$ means only the finite rational functions of the spinor variables are kept.}
\begin{eqnarray}
 \mathcal{I}^{\alpha-\beta+}_{\text{d.p.}} \biggr|_{\mathbb{Q}}\!\!\!\!\! &=& 
 -\frac{i}{18} \frac{[ab]}{\spa a.b} \frac{\la q |ab|q\ra [q|\PP_{ab} |q\ra}{s_{ab}[q|\PP_{ab} |c\ra\spa{d}.e^2 }\Biggr( -{[fc]^3\over t_{ab c}[f|\PP_{ab}|c\ra}\frac{[q|\PP_{ab}|c\ra^3}{[q|\PP_{ab}|q\ra^3}
\notag \\
 & & \quad\quad\quad\quad\quad\quad +{ [ef]\spa{d}.f \over\spa{c}.d\spa{e}.f^2}\frac{[q|\PP_{ab}|e\ra^3}{[q|\PP_{ab}|q\ra^3}
-{\spa{c}.e [dc]\over\spa{c}.d^2\spa{e}.f }\frac{[q|\PP_{ab}|d\ra^3}{[q|\PP_{ab}|q\ra^3}\frac{[q|\PP_{ab}|c\ra}{[q|\PP_{ab}|f\ra }\Biggr)\; .
\notag \\
\end{eqnarray}
The expression for $\mathcal{I}^{\alpha-\beta+}_{\text{n.f.}}$ is quite complicated and is given in (\ref{eq:intnf}).
The full $s_{ab}$ contribution requires the other helicity configuration in fig.~\ref{fig:axialg} whose contribution can be deduced by the application of 
symmetry principles to $\mathcal{I}^{\alpha-\beta+}$ after integration as detailed in appendix~\ref{app:B}.

For $\mathcal{I}^{\alpha\beta}_{\text{s.b.}}$, however, there are useful cancellations between the two helicity choices for $\alpha$ and $\beta$ before 
integration. Defining $\mathcal{I}=\mathcal{I}^{\alpha-\beta+}+\mathcal{I}^{\alpha+\beta-}$, the full contribution from the square bracket factorisation is 
\begin{align}
 \mathcal{I}_{\text{s.b.}} \biggr|_{\mathbb{Q}}\!\!\!\!\!& =  \frac{i}{9} {[qa][qb]\over \spa a.b}\frac{[f|\PP_{ab}|q\ra^2}{ [q|\PP_{ab}|q\ra^2t_{cde}}\biggl( {\la q |\PP_{ab}|c][cd]\over \spa{d}.e\la q |\PP_{ab}\PP_{cd}|e\ra} -{[de][ef]\over \spa{c}.d[f|k|c\ra}+{[ce]\over\spa{c}.d\spa{d}.e} \biggr)\Biggr]\; .
\end{align}

We can now compute $R_6^{(2)}$ by recursion.
\section{$R_6^{(2)}$}

 As discussed above, the BCFW shift does not vanish at infinity and so cannot be used to generate the result. 
At five-point this can be seen by applying the shift to the known result, while at six- and seven-point it can be seen retrospectively by applying 
the shift to the results we obtain. However the Risager shift does vanish 
at infinity for the five-point amplitude and results in rational terms with the correct symmetries and factorisations for both the six- and seven-point amplitudes. 
This self-consistency provides a stringent check of the result. 

\def\ki{a}
\def\kj{b}
\def\kk{c}

Applying the Risager shift to three adjacent legs, $\ki$, $\kj$ and $\kk$ of $R_6^{(2)}(a,b,c,d,e,f)$ and identifying $\lambda_\eta$ with the axial gauge spinor 
$\lambda_{q}$\footnote{Setting $\lambda_\eta=\lambda_q$ prevents the shift from exciting spurious poles such as $\la\hat \ki q\ra \longrightarrow 0$.}, 
factorisations arise with both single and double poles. Those involving only simple poles are
\begin{align}
&\sum_{\lambda=\pm} A^{(1)}_{4}(e,f,\hat{\ki}^{_+},\hat{\PP}^\lambda){1\over \PP^2} 
 A^{(1)}_{4}(-\hat{\PP}^{-\lambda},\hat{\kj},\hat{\kk},d)\, ,
\notag \\ 
&\sum_{\lambda=\pm}\! A^{(1)}_{4}(f,\hat{\ki},\hat{\kj},\hat{\PP}^\lambda){1\over \PP^2}  
A^{(1)}_{4}(-\hat{\PP}^{-\lambda},\hat{\kk},d,e)\quad 
\end{align}
and 
\begin{align}
\Atree_{3}(\hat{\ki},\hat{\kj},\hat{\PP}^-)&{1\over \PP^2}  R^{(2)}_{5}(-\hat{\PP}^{+},\hat{\kk},d,e,f)\; ,
\notag \\ 
\Atree_{3}(f,\hat{\ki},\hat{\PP}^-)&{1\over \PP^2}  R^{(2)}_{5}(-\hat{\PP}^{+},\hat{\kj},\hat{\kk},d,e)\;,
\notag \\ 
R^{\text{2-loop}}_{5}(-\hat{\PP}^{+},e,f,\hat{\ki},\hat{\kj})&{1\over \PP^2}\Atree_{3}(\hat{\kk},d,\hat{\PP}^-) \; ,
\notag \\ 
  R^{(2)}_{5}(-\hat{\PP}^{+},d,e,f,\hat{\ki})&{1\over \PP^2}\Atree_{3}(\hat{\kj},\hat{\kk},\hat{\PP}^-) \; , 
\notag \\ 
\end{align}
whose residues can simply be evaluated at the appropriate complex pole.  The remaining factorisations: 
\begin{align}
&\ V^{(1)}_{3}(f^+,\hat{\ki}^{_+},\hat{\PP}^+){1\over \PP^2} 
 A^{(1)}_{5}(-\hat{\PP}^{-},\hat{\kj}^+,\hat{\kk}^+,d,e)\, ,
 \notag \\ 
&\ V^{(1)}_{3}(\hat{\kk}^+,{d}^{_+},\hat{\PP}^+){1\over \PP^2} 
 A^{(1)}_{5}(-\hat{\PP}^{-},e,f,\hat{\ki}^+,\hat{\kj}^+)\, ,
\notag \\ 
&\! V^{(1)}_{3}(\hat{\ki},\hat{\kj},\hat{\PP}^\lambda){1\over \PP^2}  
A^{(1)}_{5}(-\hat{\PP}^{-\lambda},\hat{\kk}^+,d,e,f)\; ,
 \notag \\ 
&\! V^{(1)}_{3}(\hat{\kj},\hat{\kk},\hat{\PP}^\lambda){1\over \PP^2}  
A^{(1)}_{5}(-\hat{\PP}^{-\lambda},d,e,f,\hat{\ki}^+)\; ,
\end{align} 
have double poles as determined in the previous section.  

Summing the corresponding  residues gives an expression which  involves $\lambda_\eta$ and, due to the choice of shifted legs, 
cyclic symmetry is not manifest. 
However, the result 
is independent of $\lambda_{\eta}$  and has full cyclic and flip symmetry. By construction $R_6^{(2)}$
has the correct factorisations in all of the channels excited by the shift. Combining this with the cyclic symmetry ensures that it has the correct 
factorisations in all channels. In the following section we present a compact analytic form for $R_6^{(2)}$ that is manifestly independent of 
the shift and axial gauge vectors and has manifest cyclic symmetry.

\section{Final $R_6^{(2)}$ analytic form}

We can obtain a form for $R_6^{(2)}$  that is explicitly independent of $q$, has manifest cyclic symmetry and no spurious poles. 
To obtain  this form for $R_6^{(2)}$  the form generated by recursion first has the factorising 
residues extracted:
\begin{equation}
R_6^{\text{fact}} = {i\over 9}{1\over \spa 1.2\spa 2.3 \spa 3.4 \spa 4.5 \spa 5.6 \spa 6.1}({G}_6^1+{G}_6^2+{G}_6^3+{G}_6^4)
\end{equation}
where
\begin{align}
G^1_6 =  & { s_{cd} s_{df} \la f|a K_{abc}|e \ra\over \spa{f}.{e} t_{abc}} + { s_{ac} s_{dc} \la a|f K_{def}|b \ra\over \spa{a}.{b} t_{def}}\; ,
 \notag \\
G^2_6 =  & {\spb{a}.{b} \spb{f}.{e} \over \spa{a}.{b} \spa{f}.{e} }  \spa{a}.{e}^{2} \spa{f}.{b}^2  +  {1 \over 2}{\spb{a}.{f} \spb{c}.{d} \over \spa{a}.{f} \spa{c}.{d} }  \spa{a}.{c}^{2} \spa{d}.{f}^2   \; ,
\notag \\
G^3_6 =  & { s_{df} \spa{a}.{f} \spa{c}.{d} \spb{c}.{a} \spb{d}.{f} \over t_{abc}}  
\shortintertext{and}
G^4_6 =  & { \la a |b e|f \ra t_{def} \over \spa{a}.{f} } \; .
\end{align}
Then the remainder of the function is fitted:
\begin{align}
G^5_6 =   & 2s_{ac}^2+s_{eb}^2 + s_{ab} \left(-3 s_{ac} - 2 s_{ad} +6 s_{ae} +4s_{bc} + s_{bd} +2s_{be} + 4s_{bf}+7 s_{cd} - s_{ce} - s_{de} +3 s_{df}\right) \notag \\ 
&+s_{ac} \left( 2s_{ad} +3s_{ae} -2 s_{bd} - s_{be} + s_{cf} -{5 \over 2}s_{df} \right) + {3 \over 2 }s_{ad}s_{be} \notag \\
&- 8 \spa{b}.{c} \spb{c}.{d} \spa{d}.{e} \spb{e}.{b} +5 \spa{f}.{a} \spb{a}.{c} \spa{c}.{d} \spb{d}.{f}\; ,
 \notag 
\end{align}
leading to the final expression
\begin{align}
R_6^{(2)} = {i\over 9}\sum_{\text{cyclic perms}}{G_6^1+G_6^2+G_6^3+G_6^4+G_6^5\over \spa 1.2\spa 2.3 \spa 3.4 \spa 4.5 \spa 5.6 \spa 6.1}.
\end{align}
This was confirmed in an independent calculation~\cite{Badger:2016ozq}.  This is a simpler form than that of \cite{Dunbar:2016gjb}: it has been reformulated to be manifestly free of spurious singularities.

\section{The Seven-Point Rational Piece}
The seven-point rational piece can be calculated in an identical fashion. The seven-point current $\tau_7^{(1)} (\alpha^{-},\beta^{+},c^+,d^+,e^+,f^+,g^+)$
is built from the corresponding seven-point single minus amplitude~\cite{Bern:2005hs} just as the six-point current was built from the six-point amplitude. 
$R_6^{(2)}$ as determined above is also required for recursion. Defining
\begin{align}
G^1_7 =  & {\spa g.a\over t_{abc}t_{efg}} \Biggr({\spa c.d\spb e.g[d|K_{abc}|e\ra[a|K_{abc}|e\ra [c|K_{abc}|f\ra\over \spa e.f}-{\spa d.e\spb c.a [d|K_{efg}|c\ra  [g|K_{efg}|c\ra [e|K_{efg}|b\ra  \over \spa b.c}
\notag \\ &
+ { \spa e.f \spa c.d  \spb c.a\spb f.g [e|K_{efg}|a\ra [d|K_{efg}| b\ra \over \spa a.b} - {\spa b.c \spa d.e \spb e.g \spb a.b [c|K_{abc}|g\ra [d|K_{abc}| f\ra\over \spa f.g}\Biggr)\; ,
 \notag \\
G^2_7 =  & {1\over t_{abc}t_{efg}}\, s_{cd}s_{de}\spa g.a[g|K_{efg}K_{abc}|a] \; ,
\notag \\
G^3_7 =  & {1\over t_{cde}} \Biggr( s_{ce}\Biggr({s_{ef}\la c|K_{ab}K_{fga}|d\ra\over \spa c.d}-{s_{bc}\la e|K_{fg}K_{gab}|d\ra \over \spa d.e}\Biggr)+{\spa e.f \spa b.c \spb f.b[c|K_{cde}|g\ra[e|K_{cde}|a\ra\over \spa g.a}
\notag \\ &
+{\spa{b}.c [c|K_{cde}|b\ra[e|K_{cde}|a\ra[b|K_{fg}|e\ra\over \spa a.b}
+{ \spa{e}.f [e|K_{cde}|f\ra[c|K_{cde}|g\ra[f|K_{ab}|c\ra\over \spa f.g}\Biggr) \; ,
 \notag \\
G^4_7 =  & {\spb g.a\over \spa g.a} \spa g.e \spa a.e\Biggr( {\spb d.e\over \spa d.e}\spa d.g \spa d.a+{\spb e.f\over \spa e.f}\spa f.g\spa f.a \Biggr)\ \; ,
 \notag \\
G^5_7 =  & {1\over t_{cde}}\big( \spb c.e (\spa e.f \spb d.f \la c|K_{ab} K_{fga}|d\ra+\spa b.c \spb d.b \la e|K_{fg}K_{gab}|d\ra  )
\notag \\ &
\quad\quad\quad\quad\quad\quad\quad+ \spa b.c\spa e.f(2 \spa g.a \spb c.e\spb f.g \spb a.b+  \spb b.f [e|K_{ab}K_{fg}|c]\big) \; ,
 \notag \\
G^6_7 =  & {1\over\spa g.a} (\la g|fK_{bc}|a\ra t_{efg} -\la a|bK_{ef}|g\ra t_{abc}) 
 \shortintertext{and}
G^7_7 =  & s_{bf}^2-2s_{ga}^2-3s_{db}s_{df}+4s_{da}s_{dg}-6s_{ac}s_{eg}+7(s_{eb}s_{fc}+s_{ea}s_{gc})+s_{ab}s_{fg}+3s_{fa}s_{gb}
 \notag \\ &
  +s_{ce}(s_{cf}+s_{eb}-4(s_{ab}+s_{fg}+s_{ga})+5[d|K_{ga}|d\ra)
  \notag \\ &
  +4[e|bcf|e\ra-2[f|gab|f\ra+3[g|baf|g\ra+2[g|cea|g\ra  ,
\end{align}
the full function in this case is
\begin{align}
R_7^{(2)} = {i\over 9}\sum_{\text{cyclic perms}}{G_7^1+G_7^2+G_7^3+G_7^4+G_7^5+G_7^6+G_7^7\over \spa 1.2\spa 2.3 \spa 3.4 \spa 4.5 \spa 5.6 \spa 6.7 \spa 7.1}\;.
\end{align}
This expression has the full cyclic and flip symmetries  required and has all the correct factorisations and collinear limits. 
It been generated under the assumption that the shifted rational function vanishes at infinity: if this had been unjustified we would not 
have generated a function with the appropriate symmetries. 
This completes the seven-point calculation.

\section{Conclusion}

In this article we have detailed the construction of the all-plus leading in colour two-loop gluon scattering amplitude and presented the 
results for the six- and seven-point cases in straightforward analytic expressions. Our methods are based upon an understanding of the singular 
structure an on-shell amplitude satisfies and in particular we have only needed four-dimensional unitarity methods. 
These techniques, at present, are not completely rigorous but the amplitudes generated satisfy a range of stringent consistency checks.
 We 
have largely used on-shell methods but have had to augment these with some off-shell information. General theorems for the factorisation of loop 
amplitudes on complex momenta which specified the sub-leading behaviour would simplify further the process and avoid integrations. 

This particular helicity configuration remains the only 
two-loop amplitude calculated beyond four point.  Extending analytic results to the other helicity amplitudes is clearly necessary 
for phenomenology and we hope the methods here will be fruitful in this.

\section{Acknowledgements}
This work was supported by STFC grant ST/L000369/1. GRJ was 
supported by STFC grant ST/M503848/1. JHG was 
supported by the College
of Science (CoS) Doctoral Training Centre (DTC) at 
Swansea University

\appendix

\section{The six-point one-loop single minus amplitude}

\label{app:sixpoint}
The six-point single minus one-loop amplitude is~\cite{Bern:2005hs}:
\begin{align}
 A_6^{(1)}&(\alpha^-,\beta^+,c^+,d^+,e^+,f^+)
 \notag \\
&={i\over 3} \Biggl[ {[f|\PP_{\beta c}|\alpha\ra^3\over \spa{\alpha}.\beta \spa{\beta}.c\spa{d}.e^2 t_{\alpha\beta c}[f|\PP_{\alpha\beta}|c\ra}
                              +{[\beta|\PP_{cd}|\alpha\ra^3\over \spa{c}.d^2 \spa{e}.f\spa{f}.\alpha t_{\beta cd} [\beta|\PP_{cd}|e\ra}
\notag \\ &
+{[\beta f]^3\over [\alpha\beta][f\alpha] t_{cde}}\biggl( {[\beta c][cd]\over \spa{d}.e[\beta|\PP_{cd}|e\ra}
                                      -{[de][ef]\over \spa{c}.d[f|\PP_{\alpha\beta}|c\ra}
                                      +{[ce]\over\spa{c}.d\spa{d}.e}
                              \biggr)
\notag \\ &
 -{\spa{\alpha}.c^3 [\beta c]\spa{\beta}.d \over\spa{\beta}.c^2\spa{c}.d^2\spa{d}.e\spa{e}.f\spa{f}.\alpha}
+{\spa{\alpha}.e^3 [ef]\spa{d}.f \over\spa{\alpha}.\beta\spa{\beta}.c\spa{c}.d\spa{d}.e^2\spa{e}.f^2}
\notag \\ &
-{\spa{\alpha}.d^3 \spa{c}.e [d|\PP_{\beta c}|\alpha\ra \over\spa{\alpha}.\beta\spa{\beta}.c\spa{c}.d^2\spa{d}.e^2\spa{e}.f\spa{f}.\alpha}
\Biggr]\; . 
\label{sixsm}
\end{align}

We manipulate the amplitude into a form where taking the off-shell continuation exactly reproduces the factorisations 
shown in fig.\ref{fig:doublepoleB}, thus satisfying condition {\bf I}. 
As the starting point is the six-point amplitude, condition {\bf II} is automatically encoded. First we collect terms with $\spa{\alpha}.{\beta}$ and $\spb{\alpha}.{\beta}$ in the denominator as these contribute to the factorisations.
\begin{align}
A_6^{(1)}&(\alpha^-,\beta^+,c^+,d^+,e^+,f^+)
 \notag \\
=&\frac{i}{3} \Biggl[\frac{1}{\spa{\alpha}.\beta \spa{\beta}.c\spa{d}.e^2 }\Biggr( {[f|\PP_{\beta c}|\alpha\ra^3\over t_{\alpha\beta c}[f|\PP_{\alpha\beta}|c\ra}
+{\spa{\alpha}.e^3 [ef]\spa{d}.f \over\spa{c}.d\spa{e}.f^2}
-{\spa{\alpha}.d^3 \spa{c}.e [d|\PP_{\beta c}|\alpha\ra \over\spa{c}.d^2\spa{e}.f\spa{f}.\alpha}\Biggr)
\notag \\ &
+{[\beta f]^3\over [\alpha\beta][f\alpha] t_{cde}}\biggl( {[\beta c][cd]\over \spa{d}.e[\beta|\PP_{cd}|e\ra}
                                      -{[de][ef]\over \spa{c}.d[f|\PP_{\alpha\beta}|c\ra}
                                      +{[ce]\over\spa{c}.d\spa{d}.e}
                              \biggr)
\notag \\ &
                              +{[\beta|\PP_{cd}|\alpha\ra^3\over \spa{c}.d^2 \spa{e}.f\spa{f}.\alpha t_{\beta cd} [\beta|\PP_{cd}|e\ra}
 -{\spa{\alpha}.c^3 [\beta c]\spa{\beta}.d \over\spa{\beta}.c^2\spa{c}.d^2\spa{d}.e\spa{e}.f\spa{f}.\alpha}
\Biggr]\;
\end{align}

Expanding in $\spa \alpha.\beta$ to isolate the singular pieces we find 
\begin{align}
A_6^{(1)}&(\alpha^-,\beta^+,c^+,d^+,e^+,f^+)
\notag \\
=&\frac{i}{3} \Biggl[\frac{1}{\spa{\alpha}.\beta \spa{\beta}.c\spa{d}.e^2 }\Biggr( {[fc]^3\spa{c}.\alpha ^3\over t_{\alpha\beta c}[f|\PP_{\alpha\beta}|c\ra}
+{\spa{\alpha}.e^3 [ef]\spa{d}.f \over\spa{c}.d\spa{e}.f^2}
-{\spa{\alpha}.d^3 \spa{c}.e [dc]\spa{c}.\alpha\over\spa{c}.d^2\spa{e}.f\spa{f}.\alpha}\Biggr)
\notag \\ &
+\frac{1}{\spa{\beta}.c\spa{d}.e^2 }\Biggr( {[\beta f](3[f|c|\alpha\ra^2+3[f|c|\alpha\ra[f|\beta|\alpha\ra+[f|\beta|\alpha\ra^2)\over t_{\alpha\beta c}[f|\PP_{\alpha\beta}|c\ra}
-{\spa{\alpha}.d^3 \spa{c}.e [\beta d]\over\spa{c}.d^2\spa{e}.f\spa{f}.\alpha}\Biggr)
\notag \\  &
+{[\beta f]^3\over [\alpha\beta][f\alpha] t_{cde}}\biggl( {[\beta c][cd]\over \spa{d}.e[\beta|\PP_{cd}|e\ra}
                                      -{[de][ef]\over \spa{c}.d[f|\PP_{\alpha\beta}|c\ra}
                                      +{[ce]\over\spa{c}.d\spa{d}.e}
                              \biggr)
\notag \\ &
                              +{[\beta|\PP_{cd}|\alpha\ra^3\over \spa{c}.d^2 \spa{e}.f\spa{f}.\alpha t_{\beta cd} [\beta|\PP_{cd}|e\ra}
 -{\spa{\alpha}.c^3 [\beta c]\spa{\beta}.d \over\spa{\beta}.c^2\spa{c}.d^2\spa{d}.e\spa{e}.f\spa{f}.\alpha}
\Biggr]\; . 
\label{sixsm2}
\end{align}
The first line of eq.~(\ref{sixsm2}) has the $\spa \alpha.\beta ^{-1}$ factor.  
This can be rewritten using 
\begin{align}
\frac{1}{\spa \alpha.\beta\spa \beta.c} &= \frac{\spb \beta.\alpha [q|\PP_{\alpha\beta} |c\ra\spa \beta.q}{ s_{\alpha\beta}\spa \beta.c\spa \beta.q[q|\PP_{\alpha\beta} |c\ra}= \frac{\spb \beta.\alpha ([q|\PP_{\alpha\beta} |q\ra\spa \beta.c+[q|\PP_{\alpha\beta} |\beta\ra\spa c.q)}{s_{\alpha\beta}\spa \beta.c\spa \beta.q [q|\PP_{\alpha\beta} |c\ra}
\notag \\
&= \frac{1}{\spa \alpha.q \spa \beta.q ^2}\left(\frac{\la q|\alpha\beta|q\ra[q|\PP_{\alpha\beta} |q\ra}{s_{\alpha\beta}[q|\PP_{\alpha\beta} |c\ra}
+  \frac{\spa q.\beta\spa q.c[q|\alpha |q\ra}{\spa \beta.c [q|\PP_{\alpha\beta} |c\ra } \right)
\label{magicf}
\end{align}
which is an algebraic identity for any $q$, but we specifically identify $q$ with the axial gauge reference momenta. 
Applying the formula to \eqref{sixsm2}:
\begin{align}
A_6^{(1)}&(\alpha^-,\beta^+,c^+,d^+,e^+,f^+) =
\notag \\
&\frac{i}{3} \Biggl[\frac{\la q|\alpha\beta|q\ra[q|\PP_{\alpha\beta} |q\ra}{s_{\alpha\beta}[q|\PP_{\alpha\beta} |c\ra\spa \alpha.q \spa \beta.q ^2\spa{d}.e^2 }\Biggr( {[fc]^3\spa{c}.\alpha ^3\over t_{\alpha\beta c}[f|\PP_{\alpha\beta}|c\ra}
+{\spa{\alpha}.e^3 [ef]\spa{d}.f \over\spa{c}.d\spa{e}.f^2}
-{\spa{\alpha}.d^3 \spa{c}.e [dc]\spa{c}.\alpha\over\spa{c}.d^2\spa{e}.f\spa{f}.\alpha}\Biggr)
\notag \\ &
+\frac{\spa c.q[q|\alpha |q\ra}{\spa \beta.c [q|\PP_{\alpha\beta} |c\ra \spa \alpha.q \spa \beta.q \spa{d}.e^2 }\Biggr( {[fc]^3\spa{c}.\alpha ^3\over t_{\alpha\beta c}[f|\PP_{\alpha\beta}|c\ra}
+{\spa{\alpha}.e^3 [ef]\spa{d}.f \over\spa{c}.d\spa{e}.f^2}
-{\spa{\alpha}.d^3 \spa{c}.e [dc]\spa{c}.\alpha\over\spa{c}.d^2\spa{e}.f\spa{f}.\alpha}\Biggr)
\notag \\ &
+\frac{1}{\spa{\beta}.c\spa{d}.e^2 }\Biggr( {3[\beta f][f|c|\alpha\ra^2\over t_{\alpha\beta c}[f|\PP_{\alpha\beta}|c\ra}
-{\spa{\alpha}.d^3 \spa{c}.e [\beta d]\over\spa{c}.d^2\spa{e}.f\spa{f}.\alpha}\Biggr)
\notag \\  &
+{[\beta f]^3\over [\alpha\beta][f\alpha] t_{cde}}\biggl( {[\beta c][cd]\over \spa{d}.e[\beta|\PP_{cd}|e\ra}
                                      -{[de][ef]\over \spa{c}.d[f|\PP_{\alpha\beta}|c\ra}
                                      +{[ce]\over\spa{c}.d\spa{d}.e}
                              \biggr)
\notag \\ &
                              +{[\beta|\PP_{cd}|\alpha\ra^3\over \spa{c}.d^2 \spa{e}.f\spa{f}.\alpha t_{\beta cd} [\beta|\PP_{cd}|e\ra}
 -{\spa{\alpha}.c^3 [\beta c]\spa{\beta}.d \over\spa{\beta}.c^2\spa{c}.d^2\spa{d}.e\spa{e}.f\spa{f}.\alpha}+ \mathcal{O}(\spa \alpha.\beta)
\Biggr]\; . 
\label{eq:A7}
\end{align}
The prefactor in the first line of (\ref{eq:A7}) encodes the three-point vertex in diagram~\ref{fig:doublepoleB}a and the five-point single minus amplitude on the right can be reconstructed using
\begin{equation}
\frac{[\beta|\PP_{\alpha\beta}|X\ra}{[\beta|\PP_{\alpha\beta}|Y\ra} = \frac{[q|\PP_{\alpha\beta}|X\ra}{[q|\PP_{\alpha\beta}|Y\ra}+s_{\alpha\beta}\frac{\spa Y.X [\beta q]}{[\beta|\PP_{\alpha\beta}|Y\ra [q|\PP_{\alpha\beta}|Y\ra} + \mathcal{O}(s_{\alpha\beta}^2)\; ,
\label{mf2}
\end{equation}
applied to the remainder of the first line:
\begin{align}
&\;\frac{1}{\spa \alpha.q}\Biggr( {[fc]^3\spa{c}.\alpha ^3\over t_{\alpha\beta c}[f|\PP_{\alpha\beta}|c\ra}
+{\spa{\alpha}.e^3 [ef]\spa{d}.f \over\spa{c}.d\spa{e}.f^2}
-{\spa{\alpha}.d^3 \spa{c}.e [dc]\spa{c}.\alpha\over\spa{c}.d^2\spa{e}.f\spa{f}.\alpha}\Biggr)\quad\quad\quad\quad\quad\quad\quad\quad\quad
\notag \\ &
\notag \\ &
=\spa \alpha.q^2\Biggr( -{[fc]^3\over t_{\alpha\beta c}[f|\PP_{\alpha\beta}|c\ra}\frac{[\beta|\PP_{\alpha\beta}|c\ra^3}{[\beta|\PP_{\alpha\beta}|q\ra^3}
+{ [ef]\spa{d}.f \over\spa{c}.d\spa{e}.f^2}\frac{[\beta|\PP_{\alpha\beta}|e\ra^3}{[\beta|\PP_{\alpha\beta}|q\ra^3}
-{\spa{c}.e [dc]\over\spa{c}.d^2\spa{e}.f }\frac{[\beta|\PP_{\alpha\beta}|d\ra^3}{[\beta|\PP_{\alpha\beta}|q\ra^3}\frac{[\beta|\PP_{\alpha\beta}|c\ra}{[\beta|\PP_{\alpha\beta}|f\ra }\Biggr)
\notag \\ &
\notag \\ &
=\spa \alpha.q^2\Biggr( -{[fc]^3\over t_{\alpha\beta c}[f|\PP_{\alpha\beta}|c\ra}\frac{[q|\PP_{\alpha\beta}|c\ra^3}{[q|\PP_{\alpha\beta}|q\ra^3}
+{ [ef]\spa{d}.f \over\spa{c}.d\spa{e}.f^2}\frac{[q|\PP_{\alpha\beta}|e\ra^3}{[q|\PP_{\alpha\beta}|q\ra^3}
-{\spa{c}.e [dc]\over\spa{c}.d^2\spa{e}.f }\frac{[q|\PP_{\alpha\beta}|d\ra^3}{[q|\PP_{\alpha\beta}|q\ra^3}\frac{[q|\PP_{\alpha\beta}|c\ra}{[q|\PP_{\alpha\beta}|f\ra }\Biggr)
\notag \\ &
\quad\quad +\frac{3 s_{\alpha\beta}\spa \alpha.q^2[\beta q]}{[q|\PP_{\alpha\beta}|q\ra}\Biggr( -{[fc]^3\spa q.c\over t_{\alpha\beta c}[f|\PP_{\alpha\beta}|c\ra[\beta|\PP_{\alpha\beta}|q\ra }\frac{[\beta|\PP_{\alpha\beta}|c\ra^2}{[\beta|\PP_{\alpha\beta}|q\ra^2}+{[ef]\spa{d}.f\spa q.e  \over\spa{c}.d\spa{e}.f^2[\beta|\PP_{\alpha\beta}|q\ra}\frac{[\beta|\PP_{\alpha\beta}|e\ra^2}{[\beta|\PP_{\alpha\beta}|q\ra^2}
\notag \\ &
\quad\quad\quad\quad\quad\quad\quad\quad
-{ \spa{c}.e[dc] \over\spa{c}.d^2\spa{e}.f} \frac{[\beta|\PP_{\alpha\beta}|d\ra^2}{[\beta|\PP_{\alpha\beta}|f\ra[\beta|\PP_{\alpha\beta}|q\ra^2}\biggl(\spa q.d[\beta|\PP_{\alpha\beta}|c\ra+{\spa f.c [q|\PP_{\alpha\beta}|q\ra[\beta|\PP_{\alpha\beta}|d\ra\over 3[q|\PP_{\alpha\beta}|f\ra}\biggr) \Biggr) \; 
\notag \\ &
\notag \\ &
\quad\quad\quad\quad\quad\quad\quad\quad\quad\quad\quad\quad\quad\quad\quad\quad\quad\quad\quad\quad\quad\quad\quad\quad\quad\quad\quad\quad\quad\quad\quad\quad\quad\quad + \mathcal{O}(s_{\alpha\beta}^2)
\label{eq:taufunction}
\end{align}
and so we obtain 
\begin{eqnarray}
\tau^{\rm d.p}_6 & =&  \frac{i}{3} \frac{\la q|\alpha\beta|q\ra[q|\PP_{\alpha\beta} |q\ra}{s_{\alpha\beta}[q|\PP_{\alpha\beta} |c\ra\spa{d}.e^2 }\frac{\spa \alpha.q^2}{\spa \beta.q^2}\Biggr( -{[fc]^3\over t_{\alpha\beta c}[f|\PP_{\alpha\beta}|c\ra}\frac{[q|\PP_{\alpha\beta}|c\ra^3}{[q|\PP_{\alpha\beta}|q\ra^3}
 \notag \\  & &  \null \hskip 2.0truecm 
+{ [ef]\spa{d}.f \over\spa{c}.d\spa{e}.f^2}\frac{[q|\PP_{\alpha\beta}|e\ra^3}{[q|\PP_{\alpha\beta}|q\ra^3}
-{\spa{c}.e [dc]\over\spa{c}.d^2\spa{e}.f }\frac{[q|\PP_{\alpha\beta}|d\ra^3}{[q|\PP_{\alpha\beta}|q\ra^3}\frac{[q|\PP_{\alpha\beta}|c\ra}{[q|\PP_{\alpha\beta}|f\ra }\Biggr) \; .
\end{eqnarray}
When $K_{\alpha\beta}^2\rightarrow 0$ this reduces to~\footnote{From here on $k=\PP_{\alpha\beta}$; $k^\flat$ is defined as in equation \eqref{eq:nullified}.}
\begin{eqnarray}
\tau^{\rm d.p}_6 & =& \frac{\spa \alpha.q^2}{\spa \beta.q^2} {\la q|\alpha\beta|q\ra\over \spa k.q^2}\times{1\over s_{\alpha\beta}}\times A^{(1)}_5(k^-,c^+,d^+,e^+,f^+)
\end{eqnarray}
which exactly reproduces the first factorisation in fig.~\ref{fig:doublepoleB}
if we allow $\alpha$ and $\beta$ to be massive. This term  will generate the double pole after integration. 
The order $s_{\alpha\beta}^1$ term in \eqref{eq:taufunction}  will contribute to the subleading single pole in $s_{ab}$.

The fourth line in \eqref{eq:A7} contains the $\spb{\alpha}.{\beta}^{-1}$ factor. 
To extract the factorisation in fig.~\ref{fig:doublepoleB}b we use the following rearrangements:
\begin{align}
&{1\over\spb{\alpha}.\beta}\biggl( {\spb{\beta}.X\over\spb{\beta}.Y}- {\spb{k^\flat\!}.X\over\spb{k^\flat}.Y}\biggr) 
 ={1\over\spb{\alpha}.\beta} {\spb{\beta}.{k^\flat}\spb{X}.Y\over\spb{\beta}.Y\spb{k^\flat}.Y}
 =-{1\over\spb{\alpha}.\beta} {\la q|{k}|{\beta}]\spb{X}.Y\over\spb{\beta}.Y\la q|{k}|Y]}
 =-{\spa{q}.{\alpha}\spb{X}.Y\over\spb{\beta}.Y\la q|{k}|Y]}  \; , 
\label{sbid1}\\
\notag \\&
{[\beta f]^3\over [\alpha\beta][f\alpha]} 
=
{[\beta f]^3\over [\alpha\beta][f\alpha]}{\spb{\alpha}.q\over\spb{\alpha}.q}
=
{[\beta f]^2\over [\alpha\beta][f\alpha]\spb{\alpha}.q}{\spb{\alpha}.q[\beta f]}
 =
 {[\beta f]^2[q f]\over [f\alpha]\spb{\alpha}.q}
-{[\beta f]^2[\beta q]\over [\alpha\beta]\spb{\alpha}.q} \; ,
\label{sbid2}
\end{align}
and
 \begin{align}
 {[\beta f]^2\spb{\beta}.q \spa{\beta}.{\alpha} \over\spb{\alpha}.q}
 =
-{ \la\alpha k^\flat\ra\spb{\beta}.q^2\over \spb{\alpha}.q\spb{k^\flat}.q }{\spb{f}.{k^\flat}^2}
+s_{\alpha\beta} {[f|q|\alpha\ra\over [q|k|q\ra} {\bigl([f\beta]\spb{\beta}.q\spb{k^\flat}.q + \spb{\beta}.q^2\spb{f}.{k^\flat}\bigr)\over\spb{\alpha}.q\spb{k^\flat}.q}\; .
\label{sbid3}
\end{align}
Applying \eqref{sbid1} to the fourth line in \eqref{eq:A7}:
\begin{align}
&{[\beta f]^3\over [\alpha\beta][f\alpha] t_{cde}}\biggl( {[\beta c][cd]\over \spa{d}.e[\beta|\PP_{cd}|e\ra} -{[de][ef]\over \spa{c}.d[f|k|c\ra}+{[ce]\over\spa{c}.d\spa{d}.e} \biggr)
 \notag \\ &
 = {[\beta f]^3\over [\alpha\beta][f\alpha] t_{cde}}\biggl( {[k^\flat c][cd]\over \spa{d}.e[k^\flat|\PP_{cd}|e\ra} -{[de][ef]\over \spa{c}.d[f|k|c\ra}+{[ce]\over\spa{c}.d\spa{d}.e} \biggr) + \frac{\spa q.\alpha [\beta f]^3 [cd][c|\PP_{cd}|e\ra}{[f\alpha][\beta|\PP_{cd}|e\ra\la e| \PP_{cd}k|q\ra \spa d.e t_{cde}}                                      
\end{align}
Applying \eqref{sbid2} and \eqref{sbid3} to the pre-factor:
\begin{align}
{[\beta f]^3\over [\alpha\beta][f\alpha]} &= {[\beta f]^2[q f]\over[f\alpha]\spb{\alpha}.q}
+{ \la\alpha k^\flat\ra\spb{\beta}.q^2\over  s_{\alpha\beta}\spb{\alpha}.q\spb{k^\flat}.q }{\spb{f}.{k^\flat}^2}
- {[f|q|\alpha\ra\over [q|k|q\ra} {\bigl([f\beta]\spb{\beta}.q\spb{k^\flat}.q + \spb{\beta}.q^2\spb{f}.{k^\flat}\bigr)\over\spb{\alpha}.q\spb{k^\flat}.q}\; ,
\end{align}
we obtain
\begin{eqnarray}
\tau_6^{\rm s.b}  &=& 
- \frac{i}{3}
  \frac{\spa \alpha.q^2}{\spa \beta.q^2}{[q|\beta|q \ra ^2\over [q|\alpha|q\ra^2}{[q|\alpha\beta|q]\over  s_{\alpha\beta}\spb{k^\flat}.q^2 }\frac{\spb{f}.{k^\flat}^2}{t_{cde}}\biggl( {[k^\flat c][cd]\over \spa{d}.e[k^\flat|\PP_{cd}|e\ra} -{[de][ef]\over \spa{c}.d[f|k|c\ra}+{[ce]\over\spa{c}.d\spa{d}.e} \biggr)\;.
 \notag \\
\end{eqnarray}
When $K_{\alpha\beta}^2\rightarrow 0$ this reduces to
\begin{eqnarray}
\tau^{\rm s.b}_6 & =& -\frac{\spb \beta.q^2}{\spb \alpha.q^2} {[ q|\alpha\beta|q]\over \spb k.q^2}\times{1\over s_{\alpha\beta}}\times A^{(1)}_5(k^+,c^+,d^+,e^+,f^+)
\end{eqnarray}
which exactly reproduces the factorisation in fig.~\ref{fig:doublepoleB}b if we allow $\alpha$ and $\beta$ to be massive.
\linebreak

The remaining terms we gather into $\tau_6^{\rm n.f}$. Explicitly,
\begin{align}
\tau_6^{\rm n.f} = &
\frac{i}{3} \Biggl[\frac{\spa c.q[q|\alpha |q\ra}{\spa \beta.c [q|k |c\ra \spa \alpha.q \spa \beta.q \spa{d}.e^2 }\Biggr( {[fc]^3\spa{c}.\alpha ^3\over t_{\alpha\beta c}[f|k|c\ra}
+{\spa{\alpha}.e^3 [ef]\spa{d}.f \over\spa{c}.d\spa{e}.f^2}
-{\spa{\alpha}.d^3 \spa{c}.e [dc]\spa{c}.\alpha\over\spa{c}.d^2\spa{e}.f\spa{f}.\alpha}\Biggr)
\notag \\ &
+\frac{\spa \alpha.q^2}{\spa \beta.q ^2}\frac{3\la q|\alpha\beta|q\ra[\beta q]}{[\beta|k|q\ra[q|k |c\ra \spa{d}.e^2 }\Biggr( -{[fc]^3\spa q.c\over t_{\alpha\beta c}[f|k|c\ra }\frac{[\beta|k|c\ra^2}{[\beta|k|q\ra^2}+{[ef]\spa{d}.f\spa q.e  \over\spa{c}.d\spa{e}.f^2}\frac{[\beta|k|e\ra^2}{[\beta|k|q\ra^2}
\notag \\ &
\quad\quad\quad\quad\quad
-{ \spa{c}.e[dc] \over\spa{c}.d^2\spa{e}.f} \frac{[\beta|\PP_{\alpha\beta}|d\ra^2}{[\beta|\PP_{\alpha\beta}|f\ra[\beta|\PP_{\alpha\beta}|q\ra^2}\biggl(\spa q.d[\beta|\PP_{\alpha\beta}|c\ra+{\spa f.c [q|\PP_{\alpha\beta}|q\ra[\beta|\PP_{\alpha\beta}|d\ra\over 3[q|\PP_{\alpha\beta}|f\ra}\biggr)\Biggr)
\notag \\ &
+\frac{1}{\spa{\beta}.c\spa{d}.e^2 }\Biggr( {3[\beta f][f|c|\alpha\ra^2\over t_{\alpha\beta c}[f|k|c\ra}
-{\spa{\alpha}.d^3 \spa{c}.e [\beta d]\over\spa{c}.d^2\spa{e}.f\spa{f}.\alpha}\Biggr)
\notag \\  &
+{[\beta|\PP_{cd}|\alpha\ra^3\over \spa{c}.d^2 \spa{e}.f\spa{f}.\alpha t_{\beta cd} [\beta|\PP_{cd}|e\ra}
 -{\spa{\alpha}.c^3 [\beta c]\spa{\beta}.d \over\spa{\beta}.c^2\spa{c}.d^2\spa{d}.e\spa{e}.f\spa{f}.\alpha}
 \notag \\ &
+ {[\beta f]^2[q f]\over[f\alpha]\spb{\alpha}.q}
\frac{1}{t_{cde}}\biggl( {[k^\flat c][cd]\over \spa{d}.e[k^\flat|\PP_{cd}|e\ra} -{[de][ef]\over \spa{c}.d[f|k|c\ra}+{[ce]\over\spa{c}.d\spa{d}.e} \biggr)
 \notag \\ &
- {[f|q|\alpha\ra\over [q|k|q\ra} {\bigl([f\beta]\spb{\beta}.q\spb{k^\flat}.q + \spb{\beta}.q^2\spb{f}.{k^\flat}\bigr)\over\spb{\alpha}.q\spb{k^\flat}.q t_{cde}}\biggl( {[k^\flat c][cd]\over \spa{d}.e[k^\flat|\PP_{cd}|e\ra} -{[de][ef]\over \spa{c}.d[f|k|c\ra}+{[ce]\over\spa{c}.d\spa{d}.e} \biggr)
 \notag \\ &
 + \frac{\spa q.\alpha [\beta f]^3 [cd][c|\PP_{cd}|e\ra}{[f\alpha][\beta|\PP_{cd}|e\ra\la e| \PP_{cd}k|q\ra \spa d.e t_{cde}}
\Biggr]\; . 
\end{align}
Applying \eqref{mf2} and $\alpha+\beta=k$ and applying some minor rearrangements gives
\begin{eqnarray}
\tau_6^{\rm n.f}  &=& 
\frac{i}{3} \frac{\spa \alpha.q^2}{\spa \beta.q^2}\Biggl[[q|\alpha |q\ra\frac{\spa c.q}{[q|k|c\ra^2[q|k|q\ra^2 \spa{d}.e^2}\Biggr( -{[fc]^3[q|k|c\ra^3\over t_{\alpha\beta  c}[f|k|c\ra}
\notag \\ & &
+{ [ef]\spa{d}.f[q|k|e\ra^3 \over\spa{c}.d\spa{e}.f^2}
-{\spa{c}.e [dc][q|k|d\ra^3[q|k|c\ra\over\spa{c}.d^2\spa{e}.f [q|k|f\ra}\Biggr) -[c|\beta |d\ra{[q|k|c\ra \over\spa{c}.d^2\spa{d}.e\spa{e}.f[q|k|f\ra}
\notag \\  & &
-3[q|\beta |q\ra\frac{1}{[q|k |c\ra \spa{d}.e^2 }\Biggr( -{[fc]^3\spa q.c\over t_{\alpha\beta c}[f|k|c\ra }\frac{[q|k|c\ra^2}{[q|k|q\ra^2}+{[ef]\spa{d}.f\spa q.e  \over\spa{c}.d\spa{e}.f^2}\frac{[q|k|e\ra^2}{[q|k|q\ra^2}
\notag \\ & &
\quad\quad\quad\quad\quad
-{ \spa{c}.e[dc] \over\spa{c}.d^2\spa{e}.f} \frac{[q|\PP_{\alpha\beta}|d\ra^2}{[q|\PP_{\alpha\beta}|f\ra[q|\PP_{\alpha\beta}|q\ra^2}\biggl(\spa q.d[q|\PP_{\alpha\beta}|c\ra+{\spa f.c [q|\PP_{\alpha\beta}|q\ra[q|\PP_{\alpha\beta}|d\ra\over 3[q|\PP_{\alpha\beta}|f\ra}\biggr)\Biggr)
\notag \\ & &
-3[f|\beta |q\ra {[fc]^2[q|k|c\ra\over \spa{d}.e^2t_{\alpha\beta c}[f|k|c\ra[q|k|q\ra}
-[d|\beta|q\ra{[q|k|d\ra^3 \spa{c}.e \over\spa{c}.d^2\spa{d}.e^2 \spa{e}.f[q|k|c\ra[q|k|f\ra[q|k|q\ra}
\notag \\ & &
+ \Biggl( {[q|\beta|q\ra^2\over [q|\alpha|q\ra}{[f|k|q\ra \over  [q|k|q\ra^2}+{[f|\beta|q\ra [q|\beta|q\ra\over  [q|k|q\ra [q|\alpha|q\ra}+ {[f|\beta|q\ra^2\over [f|\alpha|q\ra [q|\alpha |q\ra}\Biggr) \times
 \notag \\ &  &
\quad\quad\quad\quad\quad\quad\quad\quad\quad\quad {[fq]\over  t_{cde}}\biggl( {[k^\flat c][cd]\over \spa{d}.e[k^\flat|\PP_{cd}|e\ra} -{[de][ef]\over \spa{c}.d[f|k|c\ra}+{[ce]\over\spa{c}.d\spa{d}.e} \biggr)
\notag \\ &  &
-\frac{[\beta|\PP_{cd}|\alpha\ra^3\spa \beta.q^3}{\spa{f}.\alpha t_{\beta cd} \la q|\beta \PP_{cd}|e\ra\spa \alpha.q^2}{1\over \spa{c}.d^2 \spa{e}.f}
 - \frac{[f|\beta|q\ra^3}{[f|\alpha|q\ra \la q|\beta \PP_{cd}|e\ra}\frac{ [cd][c|\PP_{cd}|e\ra}{\la e| \PP_{cd}k|q\ra \spa d.e t_{cde}}
\Biggr]\; . 
\notag \\
\label{eq:taunfneat}
\end{eqnarray}
Some of the terms can be further expanded to aid integration:
\begin{align}
[c|\beta |d\ra &= -[c|\alpha |d\ra+[c|k |d\ra= -\frac{\spa d.\alpha}{\spa q.\alpha}[c|\alpha|q\ra +[c|k |d\ra
\notag \\ 
{[q|\beta|q\ra^2\over [q|\alpha|q\ra}  &= [q|\alpha|q\ra -2[q|k|q\ra +{[q|k|q\ra^2\over [q|\alpha|q\ra}
\notag \\ 
{[f|\beta|q\ra [q|\beta|q\ra\over [q|\alpha|q\ra} &=-[f|\beta|q\ra  -{[f|\alpha|q\ra [q|k|q\ra\over [q|\alpha|q\ra}+{[f|k|q\ra [q|k|q\ra\over [q|\alpha|q\ra}
\notag \\ 
{[f|\beta|q\ra^2\over [f|\alpha|q\ra [q|\alpha |q\ra}&= {[f|\alpha|q\ra\over [q|\alpha |q\ra}-2{[f|k|q\ra \over [q|\alpha |q\ra}+{[f|k|q\ra^2\over [f|\alpha|q\ra [q|\alpha |q\ra}
\notag \\ 
\frac{[f|\beta|q\ra^3}{[f|\alpha|q\ra \la q|\beta \PP_{cd}|e\ra} &= -\frac{[f|\alpha|q\ra^2}{ \la q|\beta \PP_{cd}|e\ra}+3\frac{[f|\alpha|q\ra[f|k|q\ra}{ \la q|\beta \PP_{cd}|e\ra}-3\frac{[f|k|q\ra^2}{ \la q|\beta \PP_{cd}|e\ra}+\frac{[f|k|q\ra^3}{ [f|\alpha|q\ra \la q|\beta \PP_{cd}|e\ra}
\label{eq:intaids1}
\end{align}
The term with the $t_{\beta cd}^{-1}$ factor is the most difficult to deal with:
\begin{align}
\frac{[\beta|\PP_{cd}|\alpha\ra^3\spa \beta.q^3}{\spa{f}.\alpha t_{\beta cd} \la q|\beta \PP_{cd}|e\ra\spa \alpha.q^2} &=\frac{[\beta|\PP_{cd}|\beta\ra^3\spa \alpha.q}{\spa{f}.\alpha t_{\beta cd} \la q|\beta \PP_{cd}|e\ra}+ \mathcal{O}(\spa \alpha.\beta )
\notag \\ &\hskip -3.0 truecm
=-\frac{(t_{\beta cd}-s_{cd})^3 }{ t_{\beta cd} \la q|\beta \PP_{cd}|e\ra}\frac{[q|k|q\ra}{[q|k|f\ra} +\mathcal{O}(\spa \alpha.\beta )
\notag \\ &  \hskip -3.0 truecm
= -\frac{[q|k|q\ra}{[q|k|f\ra} \left( \frac{[\beta|\PP_{cd}|\beta\ra^2 }{\la q|\beta \PP_{cd}|e\ra}-\frac{[\beta|\PP_{cd}|\beta\ra s_{cd} }{\la q|\beta \PP_{cd}|e\ra}+{s_{cd}^2\over \la q|\beta \PP_{cd}|e\ra}-\frac{s_{cd}^3 }{ t_{\beta cd} \la q|\beta \PP_{cd}|e\ra}\right)+ \mathcal{O}(\spa \alpha.\beta )\label{eq:intaids2}
\end{align}
and
\begin{align}
[\beta|\PP_{cd}|\beta\ra &=[c|\beta|c\ra +[d|\beta|d\ra=-[c|\alpha|c\ra -[d|\alpha|d\ra+([c|k|c\ra +[d|k|d\ra)
\notag \\ &
= \frac{\la \alpha|\PP_{cd}\alpha|q\ra }{\spa q.\alpha} +([c|k|c\ra +[d|k|d\ra)\; .
\label{eq:intaids3}
\end{align}
This yields the final form for $\tau_6^{\rm n.f}$. Thus the full amplitude can be written as
\begin{equation}
A_6^{(1)}(\alpha^-,\beta^+,c^+,d^+,e^+,f^+)= \tau_6^{\rm d.p}+\tau_6^{\rm s.b}+\tau_6^{\rm n.f}+{\cal O}(\spa{\alpha}.{\beta})
\end{equation}
where
\begin{eqnarray}
\tau_6^{\rm d.p}  &=& 
 \frac{i}{3} \frac{\la q|\alpha\beta|q\ra[q|\PP_{\alpha\beta} |q\ra}{s_{\alpha\beta}[q|\PP_{\alpha\beta} |c\ra\spa{d}.e^2 }\frac{\spa \alpha.q^2}{\spa \beta.q^2}\Biggr( -{[fc]^3\over t_{\alpha\beta c}[f|\PP_{\alpha\beta}|c\ra}\frac{[q|\PP_{\alpha\beta}|c\ra^3}{[q|\PP_{\alpha\beta}|q\ra^3}
\notag \\
 & & \quad\quad\quad\quad\quad\quad\quad\quad\quad\quad\quad +{ [ef]\spa{d}.f \over\spa{c}.d\spa{e}.f^2}\frac{[q|\PP_{\alpha\beta}|e\ra^3}{[q|\PP_{\alpha\beta}|q\ra^3}
-{\spa{c}.e [dc]\over\spa{c}.d^2\spa{e}.f }\frac{[q|\PP_{\alpha\beta}|d\ra^3}{[q|\PP_{\alpha\beta}|q\ra^3}\frac{[q|\PP_{\alpha\beta}|c\ra}{[q|\PP_{\alpha\beta}|f\ra }\Biggr)
\;,
\notag \\
\tau_6^{\rm s.b}  &=& 
- \frac{i}{3}
  {[q|\beta|q \ra ^2\over [q|\alpha|q\ra^2}{[q|\alpha\beta|q]\over  [q|\PP_{\alpha\beta }|q\ra^2 }{1\over s_{\alpha\beta }}\frac{[f|\PP_{\alpha\beta }|q\ra^2}{t_{cde}}\biggl( {\la q |\PP_{\alpha\beta }|c][cd]\over \spa{d}.e\la q |\PP_{\alpha\beta }\PP_{cd}|e\ra} -{[de][ef]\over \spa{c}.d[f|k|c\ra}+{[ce]\over\spa{c}.d\spa{d}.e} \biggr)
  \notag \\
\end{eqnarray}
and
\begin{eqnarray}
\tau_6^{\rm n.f}  &=& 
\frac{i}{3} \frac{\spa \alpha.q^2}{\spa \beta.q^2}\Biggl[[q|\alpha |q\ra\frac{\spa c.q}{[q|k|c\ra^2[q|k|q\ra^2 \spa{d}.e^2}\Biggr( -{[fc]^3[q|k|c\ra^3\over t_{\alpha\beta c}[f|k|c\ra}
\notag \\ & &
+{ [ef]\spa{d}.f[q|k|e\ra^3 \over\spa{c}.d\spa{e}.f^2}
-{\spa{c}.e [dc][q|k|d\ra^3[q|k|c\ra\over\spa{c}.d^2\spa{e}.f [q|k|f\ra}\Biggr) 
\notag \\ & &
+\biggl(\frac{\spa d.\alpha}{\spa q.\alpha}[c|\alpha|q\ra -[c|k |d\ra\biggr){[q|k|c\ra \over\spa{c}.d^2\spa{d}.e\spa{e}.f[q|k|f\ra}
\notag \\  & &
-3[q|\beta |q\ra\frac{1}{[q|k |c\ra \spa{d}.e^2 }\Biggr( -{[fc]^3\spa q.c\over t_{\alpha\beta c}[f|k|c\ra }\frac{[q|k|c\ra^2}{[q|k|q\ra^2}+{[ef]\spa{d}.f\spa q.e  \over\spa{c}.d\spa{e}.f^2}\frac{[q|k|e\ra^2}{[q|k|q\ra^2}
\notag \\ & &
\quad\quad\quad\quad\quad
-{ \spa{c}.e[dc] \over\spa{c}.d^2\spa{e}.f} \frac{[q|k|d\ra^2}{[q|k|f\ra[q|k|q\ra^2}\biggl(\spa q.d[q|k|c\ra+{\spa f.c [q|k|q\ra[q|k|d\ra\over 3[q|k|f\ra}\biggr)\Biggr)
\notag \\ & &
-3[f|\beta |q\ra {[fc]^2[q|k|c\ra\over \spa{d}.e^2t_{\alpha\beta c}[f|k|c\ra[q|k|q\ra}
-[d|\beta|q\ra{[q|k|d\ra^3 \spa{c}.e \over\spa{c}.d^2\spa{d}.e^2 \spa{e}.f[q|k|c\ra[q|k|f\ra[q|k|q\ra}
 \notag \\ & &
+ \Biggl( {[f|k|q\ra \over  [q|k|q\ra^2}\biggr([q|\alpha|q\ra -2[q|k|q\ra +{[q|k|q\ra^2\over [q|\alpha|q\ra}\biggr)+ {[f|\alpha|q\ra\over [q|\alpha |q\ra}-2{[f|k|q\ra \over [q|\alpha |q\ra}+{[f|k|q\ra^2\over [f|\alpha|q\ra [q|\alpha |q\ra}
 \notag \\ & &
\quad\quad +{1\over [q|k|q\ra} \biggr(-[f|\beta|q\ra  -{[f|\alpha|q\ra [q|k|q\ra\over [q|\alpha|q\ra}+{[f|k|q\ra [q|k|q\ra\over [q|\alpha|q\ra}\biggr)\Biggr) \times
 \notag \\ &  &
\quad\quad\quad\quad\quad\quad\quad\quad\quad\quad\quad\quad\quad\quad\quad {[fq]\over  t_{cde}}\biggl( {[k^\flat c][cd]\over \spa{d}.e[k^\flat|\PP_{cd}|e\ra} -{[de][ef]\over \spa{c}.d[f|k|c\ra}+{[ce]\over\spa{c}.d\spa{d}.e} \biggr)
\notag \\ &  &
+\frac{[q|k|q\ra}{[q|k|f\ra} {1\over \spa{c}.d^2 \spa{e}.f}\biggr( \frac{(\la \alpha |\PP_{cd}\alpha|q\ra  +\spa q.\alpha([c|k|c\ra +[d|k|d\ra))^2 }{\spa q.\alpha ^2\la q|\beta \PP_{cd}|e\ra}+\frac{s_{cd}^2 }{ t_{\beta cd} \la q|\beta \PP_{cd}|e\ra}
\notag \\ &  &
\quad\quad\quad\quad\quad\quad\quad\quad\quad\quad\quad\quad\quad -\frac{s_{cd}^3 }{ t_{\beta cd} \la q|\beta \PP_{cd}|e\ra} -\frac{(\la \alpha|\PP_{cd}\alpha|q\ra  +\spa q.\alpha ([c|k|c\ra +[d|k|d\ra)) s_{cd} }{\spa q.\alpha \la q|\beta \PP_{cd}|e\ra} \biggr)
 \notag \\ &  &
 - \frac{ [cd][c|\PP_{cd}|e\ra}{\la e| \PP_{cd}k|q\ra \spa d.e t_{cde}}\times
  \notag \\ &  & 
 \quad\quad\quad\quad\quad \biggl( -\frac{[f|\alpha|q\ra^2}{ \la q|\beta \PP_{cd}|e\ra}+3\frac{[f|\alpha|q\ra[f|k|q\ra}{ \la q|\beta \PP_{cd}|e\ra}-3\frac{[f|k|q\ra^2}{ \la q|\beta \PP_{cd}|e\ra}+\frac{[f|k|q\ra^3}{ [f|\alpha|q\ra \la q|\beta \PP_{cd}|e\ra}\biggr)
\Biggr]\; , 
\notag \\
\label{eq:taunf}
\end{eqnarray}
where the expansions~(\ref{eq:intaids1}), (\ref{eq:intaids2}) and (\ref{eq:intaids3}) have been applied to (\ref{eq:taunfneat}) to produce the final form of $\tau_6^{n.f.}$. 
Using this form we take $\alpha$ and $\beta$ off-shell. The $\tau_6^{\rm d.p}$ and $\tau_6^{\rm s.b}$ terms then exactly reproduce the two contributions  
in fig.~\ref{fig:doublepoleB}, while
$\tau_6^{\rm n.f}$ is finite as $s_{{\alpha}{\beta}}\longrightarrow 0$. The first two terms then ensure that condition {\bf I} is satisfied and all 
three  reproduce the amplitude in the on-shell limit up to terms of order  $\spa{\alpha}.{\beta}$.

\section{Integrations}
\label{app:B}

In order to determine the contribution from the diagram shown in fig.~\ref{fig:axialg} we need to evaluate the integral in (\ref{eq:tauintinitA}):
\begin{equation}
 i\frac{1}{\spa a.q\spa b.q}\int\!\! \frac{d^D\ell}{\ell^2\alpha^2\beta^2}[a|\ell|q\ra [b|\ell|q\ra \frac{\spa \beta.q^2}{\spa \alpha.q^2}\;\tau_6\; .
\end{equation}
Feynman parametrisation is carried out in the usual way:
\begin{align}
\frac{1}{\ell^2\alpha^2\beta^2} &= \Gamma (3) \int_0^1\!\!\int_0^{1-x_2}\!\!\!\!\!\!\! dx_1dx_2 \frac{1}{(\ell^2 + 2\ell\cdot (x_1a-x_2b))^3}
\notag \\ &
=  \Gamma (3) \int_0^1\!\!\int_0^{1-x_2}\!\!\!\!\!\!\! dx_1dx_2 \frac{1}{((\ell+x_1a-x_2b)^2+x_1x_2s_{ab})^3}\; ;
\end{align}
where upon the shift $\ell \rightarrow \ell - x_1a+x_2b $ the denominator becomes symmetric. The numerators of $\mathcal{I}_\text{d.p.}$ and $\mathcal{I}_\text{n.f.}$ contain loop momenta contracted solely with $\lambda_q$, thus all quadratic and higher tensor reductions vanish.

 \subsection*{The $\mathcal{I}_\text{d.p.}$ piece}

Extracting the loop momentum dependent terms from $\mathcal{I}_\text{d.p.}$ we have
\begin{align}
&\int\!\! \frac{d^D\ell}{\ell^2\alpha^2\beta^2}[a|\ell|q\ra [b|\ell|q\ra\la q|\alpha \beta|q\ra \quad\quad\quad\quad\quad\quad\quad\quad\quad\quad\quad\quad\quad\quad\quad
\notag \\ &
=  \Gamma (3) \int_0^1\!\!\int_0^{1-x_2}\!\!\!\!\!\!\! dx_1dx_2 \int\!\! \frac{d^D\ell}{((\ell+x_1a-x_2b)^2+x_1x_2s_{ab})^3}[a|\ell|q\ra [b|\ell|q\ra\la q|(\ell +a) k|q\ra
\notag \\ &
=  \Gamma (3) \int_0^1\!\!\int_0^{1-x_2}\!\!\!\!\!\!\! dx_1dx_2 \int\!\! \frac{d^D\ell}{(\ell^2+x_1x_2s_{ab})^3}x_1x_2[a|b|q\ra [b|a|q\ra\la q|(x_2b+(1-x_1)a) k|q\ra
\notag \\ &
=  \Gamma (3)\frac{-2\pi^{\frac{D}{2}}i}{\Gamma(\frac{D}{2})} \int_0^1\!\!\int_0^{1-x_2}\!\!\!\!\!\!\! dx_1dx_2[a|b|q\ra [b|a|q\ra\la q|ab|q\ra x_1x_2(1-x_1-x_2) \int_0^\infty\!\! \frac{\ell^{D-1}_Ed\ell_E}{(\ell_E^2-x_1x_2s_{ab})^3}
\;.\label{eq:feynpar}
\end{align}
The integral
\begin{align}
\int_0^\infty\!\! \frac{\ell^{D-1}_Ed\ell_E}{(\ell_E^2+\Delta)^3} 
= 
\frac{1}{2}\int_0^\infty\!\! \frac{(\ell_E^2)^{\frac{D}{2}-1}d\ell^2_E}{(\ell_E^2+\Delta)^3}
=
\frac{\Delta^{\frac{D}{2}-3}}{2} \frac{\Gamma (3-\frac{D}{2})\Gamma (\frac{D}{2})}{\Gamma (3)}
\;.\end{align}
So 
\begin{align}
\int\!\! \frac{d^D\ell}{\ell^2\alpha^2\beta^2}[a|\ell|q\ra [b|\ell|q\ra\la q|\alpha \beta|q\ra  =  
&-i\pi^{\frac{D}{2}}\Gamma (3-\frac{D}{2})\frac{[ab]^3\spa a.q^2\spa b.q^2}{s_{ab}^{3-\frac{D}{2}}} 
\notag \\ &
\int_0^1\!\!\int_0^{1-x_2}\!\!\!\!\!\!\! dx_1dx_2[(1-x_1-x_2)(-x_1x_2)^{\frac{D}{2}-2} ]
\end{align}
and
\begin{align}
\int_0^1\!\!\int_0^{1-x_2}\!\!\!\!\!\!\! dx_1dx_2[(1-x_1-x_2)(-x_1x_2)^{\frac{D}{2}-2} ]&= \int_0^1\!\!\!  dx_1x_1^{\frac{D}{2}-1} (1-x_1)^{\frac{D}{2}} 
(-\frac{1}{\frac{D}{2}-1}
+\frac{2}{D})
\notag \\ &
=\frac{\Gamma (\frac{D}{2}-1)\Gamma (\frac{D}{2}+1)}{\Gamma (D)}\frac{-4}{D(D-2)}
\notag \\ &
=\frac{\Gamma (1-\epsilon)\Gamma (3-\epsilon )}{\Gamma (4-2\epsilon)}\frac{-1}{(2-\epsilon)(1-\epsilon)}
\notag \\ &
=\frac{\Gamma (1-\epsilon)^2}{\Gamma (4-2\epsilon)}(-1)\; ;
\end{align}
so that we can simply write
\begin{align}
\int d\Lambda^{\alpha\beta}_{ab}\bigl[\la q|\alpha \beta|q\ra\bigr]   =  -\frac{1}{6} \frac{[ab]}{\spa a.b}\la q |ab|q\ra+\mathcal{O}(\epsilon)\; .
\end{align}

\subsection*{The $\mathcal{I}_\text{n.f.}$ piece}

Another algebraic step is needed before integration can easily be carried out on $\tau_6^{\rm n.f.}$:
\begin{equation}
{\spa X.\alpha \over \spa Y.\alpha} = {\spa X.a\over \spa Y.a} +\mathcal{O}(\spa a.\alpha )  \; ,
\end{equation}
implies that
\begin{align}
\frac{(\la \alpha|\PP_{cd}\alpha|q\ra  +\spa q.\alpha ([c|k|c\ra +[d|k|d\ra))}{\spa q.\alpha \la q|\beta \PP_{cd}|e\ra} = \frac{(\la a|\PP_{cd}\ell|q\ra  +\spa q.a ([c|b|c\ra +[d|b|d\ra))}{\spa q.a \la q|\beta \PP_{cd}|e\ra} +\mathcal{O}(\spa a.\alpha ) .
\end{align}
After this step, the relevant triangle integrals are
\begin{align}
\int d\Lambda^{\alpha\beta}_{ab}\bigl[1\bigr]   &=  -\frac{1}{2}\frac{[ab]}{\spa a.b}+\mathcal{O}(\epsilon)\; ,
\notag \\ 
\int d\Lambda^{\alpha\beta}_{ab}\bigl[[X|\alpha|q\ra\bigr]  &=  -\frac{1}{6}\frac{[ab]}{\spa a.b}[X|b+2a|q\ra +\mathcal{O}(\epsilon)\; ,
\notag \\ 
\int d\Lambda^{\alpha\beta}_{ab}\bigl[ [X|\beta|q\ra \bigr] &=  -\frac{1}{6}\frac{[ab]}{\spa a.b}[X|2b+a|q\ra+\mathcal{O}(\epsilon)\;\;.
\end{align}
Integrals with extra propagators involve a little bit more subtlety. 

Setting $\PP_\chi =\lambda_q\bar{\lambda}_Y$, 
some terms in  (\ref{eq:taunf}) have
factors which can be promoted to full propagators:
\begin{align}
{1\over [Y|\beta|q\ra } = \frac{1}{(\beta + \PP_\chi)^2}+\mathcal{O}(\beta^2)\; .
\label{eq:promprop}
\end{align}
The numerator can be re-written using
\begin{align}
 [a|\ell|q\ra[b|\ell|q\ra 
=  & {(\alpha^2-\ell^2) \la q| b \ell|q\ra   +(\beta^2-\ell^2)\la q| a\ell|q\ra + \ell^2\la q| ba|q\ra   \over \spa{a}.b} 
\notag \\
=  & {\beta^2\la q| a\ell|q\ra +\alpha^2\la q| b \ell|q\ra    - \ell^2\la q|(\ell- b)\PP_{ab}|q\ra   \over \spa{a}.b} \; .
\label{eq:manipole}
 \end{align}
 The pole has now been made manifest, thus the integral need not contribute a pole to contribute to the residue. For each term in (\ref{eq:manipole}) the $\alpha^2$, $\beta^2$ and $\ell^2$ factors cancel with the propagators in $\int d\Lambda$.
 For terms involving a single extra loop momentum dependent factor in the denominator this yields a sum of triangle integrals.
 For the $\alpha^2$ term in  (\ref{eq:manipole})
\begin{align}
\frac{\la q| b \ell|q\ra }{\ell^2\beta^2(\beta+\PP_\chi)^2} &\rightarrow {\la q| b \ell|q\ra\over ((\beta-x_2b+x_1\chi)^2 +x_1x_2[b|\chi|b\ra)^3}
\notag \\ &
\rightarrow {\la q| b (x_2b-x_1\chi)|q\ra\over (\beta^2 +x_1x_2[b|\chi|b\ra)^3}
=0\quad ,
\end{align}
 and for the $\ell^2$ term
\begin{align}
\frac{\la q|(\ell- b)\PP_{ab}|q\ra}{\alpha^2\beta^2(\beta+\PP_\chi)^2} &\rightarrow {\la q|(\ell- b)\PP_{ab}|q\ra\over ((\beta-x_2(b+a)+x_1\chi)^2 -x_2s_{ab}+x_1x_2[\chi|\PP_{ab}|\chi\ra)^3}
\longrightarrow 0 \quad .
\end{align}
 Thus only the $\beta^2$ term survives. For the cubic box
\begin{align}
\frac{\la q| a \ell|q\ra[X|\ell|q\ra }{\ell^2\alpha^2(\beta +\PP_\chi)^2} &\rightarrow {\la q| a \ell|q\ra[X|\ell|q\ra\over ((\ell+x_2a-x_1(b+\chi))^2+x_1x_2(s_{ab}+[a|\chi|a\ra) -(x_1-x_1^2)[b|\chi|b\ra )^3}
\notag \\ &
\rightarrow {x_1\la q| a (b+\chi)|q\ra(-x_2[X|a|q\ra+x_1[X|b+\chi|q\ra)\over (\ell^2 +x_1x_2(s_{ab}+[a|\chi|a\ra) -(x_1-x_1^2)[b|\chi|b\ra)^3}
\notag \\ &
= {x_1\la q|ab|q\ra(-x_2[X|a|q\ra+x_1[X|b|q\ra)\over (\ell^2 +x_1x_2[a|\chi|a\ra -(x_1-x_1^2)[b|\chi|b\ra)^3}+\mathcal{O}(s_{ab})\; ;
\end{align} 
so in this case 
\begin{equation}
\Delta =  -x_1x_2([a|\chi|a\ra) +(x_1-x_1^2)[b|\chi|b\ra\; .
\end{equation}
Carrying out the integration:
\begin{align}
\int d\Lambda^{\alpha\beta}_{ab}\Biggl[\frac{[X|\ell|q\ra }{(\beta +\PP_\chi)^2}\Biggr]\Biggr|_{\mathbb{Q}}\! &= {1 \over 2}{[ab]\over\spa a.b}{[X|a|q\ra\over [a|\chi|a\ra }
\;.\end{align} 
Quadratic boxes and quadratic or cubic pentagons lead to nothing but transcendental functions:
\begin{align}
\int d\Lambda^{\alpha\beta}_{ab}\Biggl[\frac{1}{(\alpha+X)^2}\Biggr]\Biggr|_{\mathbb{Q}}\!  &=  0 \; ,
\notag \\
\int d\Lambda^{\alpha\beta}_{ab}\Biggl[\frac{1}{(\alpha+X)^2(\beta+Y)^2}\Biggr]\Biggr|_{\mathbb{Q}}\!  &=  0 \; ,
\notag \\
\int d\Lambda^{\alpha\beta}_{ab}\Biggl[\frac{[W|\ell|Z\ra}{(\alpha+X)^2(\beta+Y)^2}\Biggr]\Biggr|_{\mathbb{Q}}\!  &=  0 \; .
\end{align}
 The only other type of term in \eqref{eq:taunf} is
\begin{eqnarray}
\int d\Lambda^{\alpha\beta}_{ab}\Biggl[\frac{[X|\ell|q\ra^2 }{(\beta +\PP_\chi)^2}\Biggr]\Biggr|_{\mathbb{Q}}\! &=&- {1\over 6}{\spb a.b\over \spa a.b} {[X|a|q\ra \over   [a|\chi|a\ra}  \Biggr([X|a|q\ra \biggr(1 - 2{[b|\chi|b\ra \over [a|\chi |a\ra } \biggr)-2[X|b|q\ra \Biggr)  \; .
\notag \\
\end{eqnarray}
The integrated non factorising contribution is thus (with $k=\PP_{ab}$)
\begin{eqnarray}
\mathcal{I}^{\alpha-\beta+}_{\text{n.f.}} & &= -{i\over 6} {\spb a.b \over \spa a.b} \Biggl[{1\over 3}[q|b+2a|q\ra\frac{\spa c.q}{[q|k|c\ra^2[q|k|q\ra^2 \spa{d}.e^2}\times
\notag \\ & &
\quad\quad\quad\quad\quad\quad\quad\quad\quad\quad\Biggr( -{[fc]^3[q|k|c\ra^3\over t_{abc}[f|k|c\ra}+{ [ef]\spa{d}.f[q|k|e\ra^3 \over\spa{c}.d\spa{e}.f^2}
-{\spa{c}.e [dc][q|k|d\ra^3[q|k|c\ra\over\spa{c}.d^2\spa{e}.f [q|k|f\ra}\Biggr) 
\notag \\ & &
+\biggl({1\over 3}\frac{\spa d.a}{\spa q.a} [c|b+2a|q\ra -[c|k |d\ra\biggr){[q|k|c\ra \over\spa{c}.d^2\spa{d}.e\spa{e}.f[q|k|f\ra}
\notag \\  & &
-[q|2b+a|q\ra\frac{1}{[q|k |c\ra \spa{d}.e^2 }\times
\notag \\ & &
\quad\quad
\Biggr( -{[fc]^3\spa q.c\over t_{\alpha\beta c}[f|k|c\ra }\frac{[q|k|c\ra^2}{[q|k|q\ra^2}+{[ef]\spa{d}.f\spa q.e  \over\spa{c}.d\spa{e}.f^2}\frac{[q|k|e\ra^2}{[q|k|q\ra^2}
\notag \\ & &
\quad\quad\quad\quad\quad\quad\quad\quad\quad\quad
-{ \spa{c}.e[dc] \over\spa{c}.d^2\spa{e}.f} \frac{[q|k|d\ra^2}{[q|k|f\ra[q|k|q\ra^2}\biggl(\spa q.d[q|k|c\ra+{\spa f.c [q|k|q\ra[q|k|d\ra\over 3[q|k|f\ra}\biggr)\Biggr)
\notag \\ & &
-[f|2b+a|q\ra {[fc]^2[q|k|c\ra\over \spa{d}.e^2t_{abc}[f|k|c\ra[q|k|q\ra}
-{1\over 3}[d|2b+a|q\ra{[q|k|d\ra^3 \spa{c}.e \over\spa{c}.d^2\spa{d}.e^2 \spa{e}.f[q|k|c\ra[q|k|f\ra[q|k|q\ra}
\notag \\ & &
+\Biggl( {[f|k|q\ra \over  [q|k|q\ra^2}\biggr({1\over 3} [q|b+2a|q\ra -2[q|k|q\ra \biggr) -{1\over 3}{[f|2b+a|q\ra\over [q|k|q\ra} \Biggr) \times
 \notag \\ &  &
\quad\quad\quad\quad\quad\quad\quad\quad\quad\quad\quad\quad\quad\quad\quad {[fq]\over  t_{cde}}\biggl( {\la q|k| c][cd]\over \spa{d}.e\la q| k|\PP_{cd}|e\ra} -{[de][ef]\over \spa{c}.d[f|k|c\ra}+{[ce]\over\spa{c}.d\spa{d}.e} \biggr)
\notag \\ &  &
+\frac{[q|k|q\ra}{[q|k|f\ra} {1\over \spa{c}.d^2 \spa{e}.f}\Biggr({1\over 3}{1\over \spa q.a^2}\frac{\la a|\PP_{cd}a |q\ra}{\la q|a\PP_{cd}|e\ra}\biggr(\la a|\PP_{cd}a |q\ra\biggr( 1-2 {\la q|b\PP_{cd}|e\ra\over \la q|a\PP_{cd}|e\ra}\biggr)-2\la a|\PP_{cd}b |q\ra \biggr)
\notag \\ &  &
\quad\quad\quad\quad\quad\quad\quad\quad\quad\quad\quad\quad\quad -{\la a|\PP_{cd} a |q\ra\over\la q|a\PP_{cd}|e\ra}\biggl(\frac{2([c|b|c\ra +[d|b|d\ra)- s_{cd} }{\spa q.a}\biggr) \Biggr)
 \notag \\ &  &
 +\frac{ [cd][c|\PP_{cd}|e\ra}{\la e| \PP_{cd}k|q\ra \spa d.e t_{cde}}\times
  \notag \\ &  & 
 \Biggl({1\over 3} { [f|a|q\ra \over  \la q|a \PP_{cd}|e\ra} \Biggr( [f|a|q\ra\biggr(1- 2{\la q|b \PP_{cd}|e\ra \over \la q|a \PP_{cd}|e\ra}\biggr)\! -\!2[f|b|q\ra\Biggr)\!    -\!(2[f|a|q\ra\! -\! 3[f|k|q\ra){[f|a|q\ra\over \la q|a \PP_{cd}|e\ra} \biggr) \Biggr)
 \Biggr]\; . 
\notag \\
\label{eq:intnf}
\end{eqnarray}

\subsection*{The $\mathcal{I}_{\text{s.b.}}$ piece}
Finally, for the square bracket term:
\begin{align}
\mathcal{I}^{\alpha-\beta+}_{\text{s.b.}}&= \int d\Lambda^{\alpha\beta}_{ab}\Biggl[-{i\over 3}{[q|\beta|q \ra ^2\over [q|\alpha|q\ra^2}{[q|\alpha\beta|q]\over  s_{\alpha\beta}\spb{k^\flat}.q^2 }\frac{\spb{f}.{k^\flat}^2}{t_{cde}}\biggl( {[k^\flat c][cd]\over \spa{d}.e[k^\flat|\PP_{cd}|e\ra} -{[de][ef]\over \spa{c}.d[f|k|c\ra}+{[ce]\over\spa{c}.d\spa{d}.e} \biggr)\Biggr]
\notag \\ &
=  \int d\Lambda^{\alpha\beta}_{ab}\Biggl[-\frac{\spa \alpha.q^2}{\spa \beta.q^2}{[q|\beta|q \ra ^2\over [q|\alpha|q\ra^2}{[q|\alpha\beta|q]\over  \spb{k^\flat}.q^2 }{1\over s_{ab}} \Aloop_5(k^{\flat+},c^+,d^+,e^+,f^+)\Biggr]\;.
\end{align}
In this form this term can be summed with the other internal helicity configuration to give:
\begin{align}
\mathcal{I}_{\text{s.b.}}=\mathcal{I}^{\alpha-\beta+}_{\text{s.b.}}+\mathcal{I}^{\alpha+\beta-}_{\text{s.b.}}&= \int d\Lambda^{\alpha\beta}_{ab}\Biggl[-\left( {[q|\beta|q \ra ^2\over [q|\alpha|q\ra^2}+{[q|\alpha|q \ra ^2\over [q|\beta|q\ra^2}\right){[q|\alpha\beta|q]\over  \spb{k^\flat}.q^2 }{1\over s_{ab}} \Aloop_5(k^{\flat+},c^+,d^+,e^+,f^+)\Biggr]\;.
\end{align}
Using
\begin{align}
& {[q|\beta|q \ra ^2\over [q|\alpha|q\ra^2} =  1-2 {[q|k|q \ra \over [q|\alpha|q\ra}+ {[q|k|q \ra ^2\over [q|\alpha|q\ra^2}
 \notag \\ &
 {[q|\alpha|q \ra ^2\over [q|\beta|q\ra^2} =  1-2 {[q|k|q \ra \over [q|\beta|q\ra}+ {[q|k|q \ra ^2\over [q|\beta|q\ra^2}\; ,
\end{align}
 as the final term in each case corresponds to a cubic pentagon and
\begin{equation}
{[q|k|q \ra \over [q|\alpha|q\ra}+{[q|k|q \ra \over [q|\beta|q\ra} = {[q|k|q \ra^2 \over [q|\alpha|q\ra[q|\beta|q\ra} \;
\end{equation}
also leads to a cubic pentagon only the first term contributes:
\begin{align}
\mathcal{I}_{\text{s.b.}}&= \int d\Lambda^{\alpha\beta}_{ab}\Biggr[-2{[q|\alpha\beta|q]\over  \spb{k^\flat}.q^2 }{1\over s_{ab}} \Aloop_5(k^{\flat+},c^+,d^+,e^+,f^+)\Biggr]+{\rm transcendental \;functions}\;.
\end{align}
There is the obvious problem that quadratic terms in the numerator will be present, as the loop momenta are not adjacent to a $\lambda_q$:
\begin{align}
[a|\ell|q\ra [b|\ell|q\ra[q|\alpha\beta|q]&\rightarrow [a|\ell+x_2b|q\ra [b|\ell-x_1a|q\ra [q|(\ell+(1-x_1)a+x_2b) K|q]
\notag \\
&\rightarrow {\ell^2\over 4} (([aq]+[bq])[q|K|q\ra +x_1x_2[ab]^2\spa a.q\spa b.q) (1-x_1-x_2) [q|ab|q] \; .
\end{align}
The logarithmically divergent integral is captured by the cut-constructible piece and thus dropped. The remaining rational integral leads to
\begin{align}
\mathcal{I}_{\text{s.b.}}\Biggr|_{\mathbb{Q}}\!  &= \frac{1}{3} {1\over \spa a.b}{[qa][qb]\over  \spb{k^\flat}.q^2 } \Aloop_5(k^{\flat+},c^+,d^+,e^+,f^+) \; ,
\end{align}
which corresponds to the rational part of the one-loop $(+,+;-)$ splitting function \cite{Bern:1995ix}. 

$\mathcal{I}^{\alpha+\beta-}_{\text{d.p.}}$ and $\mathcal{I}^{\alpha+\beta-}_{\text{n.f.}}$ can be obtained from  $\mathcal{I}^{\alpha-\beta+}$ 
using the reflection symmetry of the current 
\begin{align}
\mathcal{I}^{\alpha+\beta-}(a,b,c,d,e,f;q)&=\mathcal{I}^{\alpha-\beta+}(b,a,f,e,d,c;q) \;.
\end{align}
The final expression:
\begin{equation}
\mathcal{I}(a^+,b^+,c^+,d^+,e^+,f^+,g^+;q)= \mathcal{I}_{\text{d.p.}} +\mathcal{I}_{\text{s.b.}}+\mathcal{I}_{\text{n.f.}}
\end{equation}
can then be divided by $z$ and shifted to give the residues.

\end{document}